\documentclass[nofootinbib, twocolumn,superscriptaddress,preprintnumbers,letterpaper,natbib,nofootinbib]{revtex4-1}
\pdfoutput = 1
\usepackage{amssymb}
\usepackage{amsmath}
\usepackage{epsfig}
\usepackage[usenames,dvipsnames]{xcolor}
\usepackage[breaklinks,colorlinks,urlcolor=blue,citecolor=blue,linkcolor=magenta]{hyperref}
\usepackage{cleveref}
\usepackage{multirow}
\usepackage{capt-of}
\usepackage{siunitx}
\usepackage{graphicx}
\usepackage{slashed}
\usepackage{tabularx}
\usepackage{multirow}
\usepackage{braket}
\usepackage{amsfonts}
\usepackage{mathtools}
\usepackage{mathrsfs}
\usepackage[compat=1.1.0]{tikz-feynman}
\usepackage{subfiles}
\usepackage{slashed}
  
\usepackage[normalem]{ulem}
\usepackage[shortlabels]{enumitem}  

\usepackage{fontawesome5}

\definecolor{color_git}{rgb}{0.098, 0.160, 0.345}
\newcommand{\gitlink}{\href{https://github.com/stefanmarinus/CLASS_neutrinophilic}{\textsc{g}it\textsc{h}ub {\large\color{color_git}\faGithub}}}

\definecolor{orcidgreen}{HTML}{A6CE39}
\newcommand{\orcid}[1]{\href{http://orcid.org/#1}{ {\color{orcidgreen}\faOrcid} }}

\makeatletter

\begin{document}

\title{
Precision CMB constraints on eV-scale bosons coupled to neutrinos
}

\author{Stefan Sandner  \orcid{0000-0002-1802-9018} }
\email{stefan.sandner@ific.uv.es}
\affiliation{Instituto de F\'{\i}sica Corpuscular, Universidad de Valencia and CSIC, 
 Edificio Institutos Investigaci\'on, Catedr\'atico Jos\'e Beltr\'an 2, 46980 Spain}

\author{Miguel Escudero \orcid{0000-0002-4487-8742}}
\email{miguel.escudero@cern.ch}
\affiliation{Theoretical Physics Department, CERN, 1211 Geneva 23, Switzerland}

\author{Samuel J. Witte \orcid{0000-0003-4649-3085}}
\email{switte@icc.ub.edu}
\affiliation{Gravitation Astroparticle Physics Amsterdam (GRAPPA), Institute for Theoretical Physics Amsterdam and Delta Institute for Theoretical Physics, University of Amsterdam, Science Park 904, 1098 XH Amsterdam, The Netherlands}
\affiliation{Departament de F\'{i}sica Qu\`{a}ntica i Astrof\'{i}sica and Institut de Ciencies del Cosmos, Universitat de Barcelona, Diagonal 647, E-08028 Barcelona, Spain}

\begin{abstract} 
\noindent 
The cosmic microwave background (CMB) has proven to be an invaluable tool for studying the properties and interactions of neutrinos, providing insight not only into the sum of neutrino masses but also the free streaming nature of neutrinos prior to recombination. The CMB is a particularly powerful probe of new eV-scale bosons interacting with neutrinos, as these particles can thermalize with neutrinos via the inverse decay process, $\nu\bar{\nu} \rightarrow X$, and  suppress neutrino free streaming near recombination -- even for couplings as small as $\lambda_\nu \sim \mathcal{O}(10^{-13})$. Here, we revisit CMB constraints on such bosons, improving upon a number of approximations previously adopted in the literature and generalizing the constraints to a broader class of models.
This includes scenarios in which the boson is either spin-$0$ or spin-$1$, the number of interacting neutrinos is either $N_{\rm int} = 1,2 $ or $3$, and the case in which a primordial abundance of the species is present. We apply these bounds to well-motivated models, such as the singlet majoron model or a light $U(1)_{L_\mu-L_\tau}$ gauge boson, and find that they represent the leading constraints for masses $m_X\sim 1\, {\rm eV}$. 
Finally, we revisit the extent to which neutrino-philic bosons can ameliorate the Hubble tension, and find that recent improvements in the understanding of how such bosons damp neutrino free streaming reduces the previously found success of this proposal. 
\end{abstract}

\preprint{CERN-TH-2023-073}
\preprint{IFIC/23-13}
\preprint{FTUV-23-0413.0599}
  
\maketitle

\setlength\parskip{4pt}

\section{Introduction}

Neutrinos always comprise a sizable fraction of the energy density in the Universe. In particular, prior to matter-radiation equality they represent $\sim 40\%$ of the energy budget. Neutrinos are also the only species with a sizable anisotropic stress -- a consequence of their decoupling from the thermal plasma at $T\sim 2\,{\rm MeV}$. Collectively, these facts imply that neutrino free streaming plays an important role in the evolution of the gravitational potentials responsible for sourcing the CMB anisotropies~\cite{Bashinsky:2003tk,Chacko:2003dt,Hannestad:2004qu}. 
Current observations of the CMB by the Planck satellite~\cite{planck,Planck:2013pxb,Planck:2015fie} are compatible with the standard picture in which neutrinos are free streaming at redshifts $2000 \lesssim z \lesssim 10^5$~\cite{Taule:2022jrz} (corresponding to temperatures $ 0.5 \,{\rm eV} \lesssim T_\gamma \lesssim 25\,{\rm eV}$), implying these observations can be used to stringently constrain the existence of new light particles coupled to the neutrino sector. 

The impact of exotic neutrino interactions in cosmology, and in particular in the CMB, have been studied in various contexts, including scenarios in which: neutrinos have self-interactions that arise from heavy mediators~\cite{Cyr-Racine:2013jua,Oldengott:2014qra,Lancaster:2017ksf,Oldengott:2017fhy,Kreisch:2019yzn,Park:2019ibn,Das:2020xke,RoyChoudhury:2020dmd,Brinckmann:2020bcn,Kreisch:2022zxp, RoyChoudhury:2022rva}, neutrinos annihilate into massless scalars~\cite{Beacom:2004yd,Hannestad:2004qu,Bell:2005dr,Archidiacono:2013dua,Forastieri:2015paa,Forastieri:2019cuf,Venzor:2022hql}, neutrinos decay into light particles~\cite{Hannestad:2005ex,Basboll:2008fx,Escudero:2019gfk,Chacko:2019nej,Chacko:2020hmh,Barenboim:2020vrr,Chen:2022idm,FrancoAbellan:2021hdb}, and neutrinos temporarily thermalize with $\rm{eV}-$scale neutrino-philic scalars~\cite{Chacko:2003dt,Escudero:2019gvw,Escudero:2021rfi,EscuderoAbenza:2020egd}. 
The latter scenario is particularly interesting, as particles at the eV mass-scale can arise naturally in theories which explain the origin of neutrino masses (e.g. the majoron model)~\cite{Chikashige:1980ui,Schechter:1981cv,Akhmedov:1992hi,Rothstein:1992rh} or in weakly coupled realizations of spontaneously broken gauge flavor symmetries~\cite{He:1991qd,He:1990pn,Williams:2011qb,Escudero:2019gzq}. 
Furthermore, it has been shown that $\rm{eV}-$scale neutrino-philic scalars like the majoron could play an important role in helping to ameliorate the largest outstanding discrepancy in cosmology, the Hubble tension~\cite{Escudero:2019gvw,EscuderoAbenza:2020egd,Escudero:2021rfi} (see e.g.~\cite{Schoneberg:2021qvd,DiValentino:2021izs} for recent reviews on the Hubble tension and proposed solutions).  
However, this scenario is challenging to model, as the light bosons and neutrinos undergo an out-of-equilibrium thermalization followed by an out-of-equilibrium decay, leading to a non-trivial modification of the expansion history of the Universe.

\begin{figure*}[t]
\hspace{-0.2cm}
\includegraphics[width=1.0\textwidth]{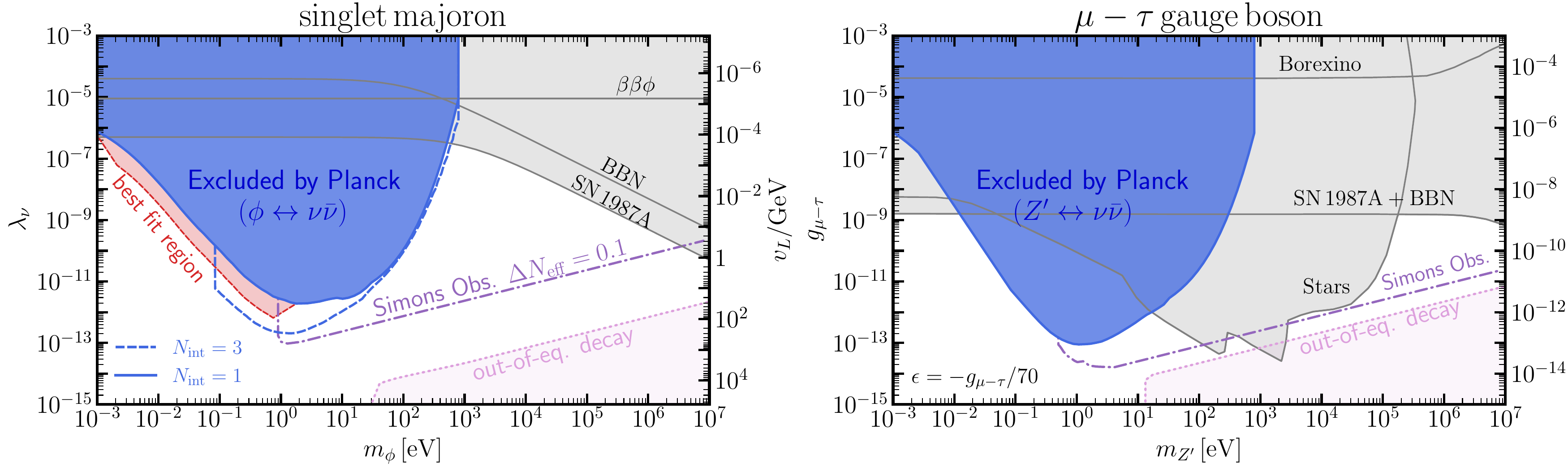}
\vspace{-0.4cm}
\caption{
Parameter space for neutrino interactions with a scalar (\textit{left panel}) and vector (\textit{right panel}) boson $X$ with mass $m_X$. The bounds are interpreted within the singlet majoron model, where $\lambda_\nu = m_\nu/v_L$ and for a light $U(1)_{L_\mu-L_\tau}$ gauge boson, for which $\lambda_\nu \simeq g_{\mu-\tau}$, respectively. 
An analysis of Planck legacy data excludes blue regions with $3\sigma$ confidence. 
Grey regions represent current cosmological, astrophysical and laboratory constraints, see Section~\ref{sec:otherconstraints} for details.
In pink we indicate constraints coming from the out-of-equilibrium decay of the new $X$ boson which apply if a primordial abundance was generated before BBN.
We also indicate the region of parameter space which will be tested by the Simons Observatory.
In particular, the region above the purple dashed-dotted line will be tested because the thermalization of the $X$ boson leads to an observable excess of $\Delta N_{\rm{eff}} \geq 0.1$.
Finally, we also highlight in red the best fit region of parameter space for the scenario of the $X$ boson being of scalar type and interacting with one neutrino family, $N_{\rm int} = 1$.
This region is of particular interest because it indicates that non-trivial neutrino interactions are statistically slightly preferred over $\Lambda$CDM. 
}
\label{fig:bounds_majoron}
\end{figure*}

The goal of this work is to perform a precision study of the impact of $\rm{eV}-$scale neutrino-philic bosons on the CMB, improving upon previous analyses which relied on numerous simplified approximations~\cite{Escudero:2019gvw,EscuderoAbenza:2020egd,Escudero:2021rfi}, and extending the results of these analyses to the more general class of light neutrino-philic bosons. The primary improvements of this work are three-fold. 
First, we have incorporated the background thermodynamic evolution of neutrinos and the neutrino-philic bosons in the cosmological Boltzmann code CLASS~\cite{Lesgourgues:2011re,Blas:2011rf}. 
This allows us to solve for the thermodynamics on the fly, with precision and speed which allows a full Bayesian analysis of Planck legacy data\footnote{
Our modified version of CLASS is available on \gitlink.
The equations for the evolution of the temperature and chemical potentials should be easily generalizable to  other scenarios involving Beyond the Standard Model (BSM) physics.}. 
Next, we incorporated a refined computation of the collision term~\cite{Barenboim:2020vrr,Chen:2022idm} which damps the neutrino free streaming less efficiently than assumed in previous studies~\cite{Escudero:2019gvw,EscuderoAbenza:2020egd,Escudero:2021rfi}. 
Finally, we generalize the analysis to arbitrary number of interacting neutrino species, include the possibility of both vector and scalar bosons and the possibility of having a primordial abundance such bosons.

In general, we find that the CMB can robustly constrain the existence of $\rm{eV}-$scale neutrino-philic bosons with couplings on the order of $\lambda_\nu \sim \mathcal{O}(10^{-13})$. 
The value of this coupling roughly corresponds to the new bosonic particles having a lifetime shorter than the age of the Universe at recombination, $\Gamma_{X} \sim \lambda_\nu^2 m_X/(8\pi) \lesssim H(z_{\rm rec})$. These bounds play an important role in testing a variety of well-motivated high-energy theories, such as the singlet majoron model (where these observations are testing scales of lepton number breaking as high as $\sim 1\,{\rm TeV}$), and the $U(1)_{L_\mu-L_\tau}$ extension of the Standard Model.  
The main results of our study are highlighted in Figure~\ref{fig:bounds_majoron}, which display the $3\sigma$ constraint on the coupling of the majoron and $U(1)_{L_\mu-L_\tau}$ gauge boson, respectively. In the case of the majoron, we also highlight a region of parameter space that is favoured by Planck legacy data at the $\sim 1\sigma$ level.

The reminder of this work is structured as follows. First, in Section~\ref{sec:models} we briefly introduce and motivate the particle physics models that we consider.  In Section~\ref{sec:formalism}, we present the formalism behind our work. In particular, we describe how we treat the thermodynamic evolution of the Universe in the presence of $\rm{eV}-$scale neutrino-philic bosons, including how the dynamics are  implemented at the level of both the background and perturbations.  In Section~\ref{sec:results} we present the constraints we derive  on the couplings between neutrinos and $\rm{eV}-$scale bosons. We also include a quantitative discussion about the ability of these models to solve or ameliorate the Hubble tension, showing that the new collision term strongly suppresses the previous success of this model identified in~\cite{Escudero:2019gvw,EscuderoAbenza:2020egd,Escudero:2021rfi}. 
Finally, in Section~\ref{sec:conclusions} we present a summary of our results and outline our conclusions. 
For completeness, we provide in the appendices~\ref{SM:theory} and~\ref{SM:cosmo} further information on the formalism and details on the modified cosmological history.

\section{Particle Physics Models}\label{sec:models}

\emph{Effective Interactions:} We will consider an effective coupling between neutrinos and a light bosonic mediator $X$ and we will study two cases, one where the mediator is a pseudoscalar $X = \phi$ and one where it is a vector $X = Z'$. We will work after electroweak symmetry breaking and in the active neutrino mass basis. The effective Lagrangians describing these interactions are:
\begin{align}
    \mathcal{L}_{\rm scalar} &= i \frac{a}{2} \,\sum_\nu \lambda_\nu \,\bar{\nu}  \gamma_5 \nu \, X \,,\label{eq:Lag_scalar} \\
    \mathcal{L}_{\rm vector} &= \frac{\sqrt{3}a}{2}\,\sum_\nu \lambda_\nu \,\bar{\nu}  \gamma^\mu P_L \nu \, X_\mu \,, \label{eq:Lag_vector} 
\end{align}
where $\lambda_\nu$ are dimensionless coupling constants and where $a = 1$ for Majorana neutrinos and $a=\sqrt{2}$ for Dirac neutrinos. 

Given these interactions, the scalar and vector boson partial decay rate into a pair of massive neutrinos are given by:
\begin{align}
\!\!\!\!    \Gamma(X \to \bar{\nu}\nu)|_{\rm scalar} &= \frac{\lambda_\nu^2 }{16\pi} m_X \sqrt{1-\frac{4m_\nu^2}{m_X^2}}  \,,  \label{eq:rates_scalar}\\
\!\!\!\!    \Gamma(X \to \bar{\nu}\nu)|_{\rm vector} &= \frac{\lambda_\nu^2 }{16\pi} m_X \sqrt{1-\frac{4m_\nu^2}{m_X^2}} \left[1-\frac{m_\nu^2}{m_X^2}\right]^{2}  \,,  \label{eq:rates_vector}
\end{align}

\emph{Mapping to concrete models:}  
These effective Lagrangians have a direct interpretation in terms of well motivated BSM scenarios. For example, Eq.~\eqref{eq:Lag_scalar} is the effective interaction generated in the famous singlet majoron model~\cite{Chikashige:1980ui} with $X = \phi$ identified as the majoron and with $\lambda_\nu = m_\nu/v_L$, where $v_L$ is the scale at which the global $U(1)_L$ symmetry is spontaneously broken.
In particular, in this model the coupling between massive neutrinos and the majoron is diagonal up to small corrections~\cite{Schechter:1981cv}. 
The vector interactions in Eq.~\eqref{eq:Lag_vector} also effectively describe new interactions of neutrinos in many BSM constructions. 
Typically, in the vector case the interaction arises by the gauging of lepton number family symmetries, and as such, the interaction is non-diagonal in the neutrino mass basis~\cite{He:1990pn,He:1991qd}.
However, in such cases all massive neutrinos couple to the $X$ boson, and the couplings in the mass and flavor basis are simply related by a PMNS rotation. As an example, we can consider the case of a light $U(1)_{L_\mu-L_\tau}$ gauge boson; 
here, the coupling $\lambda_\nu$ is intimately related to the $U(1)_{L_\mu-L_\tau}$ gauge coupling, $\lambda_\nu \simeq g_{\mu-\tau}$ -- see Ref.~\cite{Escudero:2020ped} for the precise mapping.

\emph{Scenarios Considered:} 
We will consider several scenarios that we expect to broadly cover the phenomenology of the most well-motivated BSM models featuring new neutrino interactions below the MeV scale (these scenarios are summarized in Table~\ref{tab:cases}). 

All scenarios correspond to different combinations of i) the number of interacting neutrino families, $N_{\rm int}$, ii) the internal degrees of freedom of the $X$ particle, $g_X$, and iii) if the $X$ species has a non-zero primordial abundance or not, parametrized by $\Delta N_{\rm eff}^{\rm BBN}$.
To be specific, we consider the following:
\begin{itemize}
    \item Case (a), with $N_{\rm int}=3$ and $g_X = 1$, corresponds to the singlet majoron model in which neutrinos are pseudo-degenerate (note that pseudo-degenerate neutrinos imply a universal coupling $\lambda_\nu$).
    \item Case (b), with $N_{\rm int}=3$ and $g_X = 3$,  corresponds to the commonly studied model of a light $Z'$ boson coupled to a lepton number family symmetry. In this model it is once again a good approximation to consider a flavour universal coupling, since the PMNS matrix does not show a hierarchical structure.
    \item Case (c), with $N_{\rm int}=1$ and $g_X = 1$,  corresponds to the case of the singlet majoron model coupled mainly to one neutrino. This can happen with one approximate vanishing neutrino mass eigenstate where the coupling is mostly to the heaviest neutrino state or for $ 2m_\nu^{\rm lightest}< m_X <0.1\,{\rm eV}\simeq 2\sqrt{|\Delta m_{\rm atm}^2|}$ since the majoron in that case can only kinematically couple to the lightest neutrino. 
    \item Case (d) corresponds to a case where a vector boson couples to a single neutrino mass eigenstate.  As in scenario (c), this option is relevant in particular for $2m_\nu^{\rm lightest} < m_X< 0.1\,{\rm eV}$. However, a concrete model realization for $m_X > 0.1\,{\rm eV}$ in which a vector interacts only with one neutrino mass eigenstate is challenging, and generically involves cancellations of different couplings in flavour space.
    \item The cases (e) and (f) correspond to the cases (a) and (b), respectively, but allowing for a non-zero primordial abundance of the $X$ particle parameterized by $\Delta N_{\rm eff}^{\rm BBN}$. Such a primordial abundance of $X$ particles can arise e.g. due to the decay of other, heavy particle species in the early Universe. For instance, majorons can be produced from the decays of $\rm{GeV}-$scale sterile neutrinos~\cite{Escudero:2021rfi}, and the $U(1)_{L_\mu-L_\tau}$ gauge boson can be produced via muon-antimuon annihilations in the early Universe~\cite{Escudero:2019gzq}.
\end{itemize}
\begin{table}[t]
    \centering
    \begin{tabular}{c|c} \hline \hline
       $\,\,$   Scenario $\,\,$ & $\qquad \qquad \quad$ Specification $\qquad \qquad \quad$ \\ \hline \hline
        (a)  & $N_{\rm int} = 3$, \,  $g_X = 1$ \\  
        \hline 
        (b) & $N_{\rm int} = 3$, \,  $g_X = 3$ \\  
        \hline 
        (c) & $N_{\rm int} = 1$, \, $g_X = 1$ \\  
        \hline 
        (d) & $N_{\rm int} = 1$, \, $g_X = 3$ \\  
        \hline 
        \\[-0.9em]
        (e) & $N_{\rm int} = 3$, \, $g_X = 1$, \, $\Delta N_{\rm eff}^{\rm BBN}\neq 0$ \\  
        \\[-0.9em]
        \hline 
        \\[-0.9em]
        (f) & $N_{\rm int} = 3$, \, $g_X = 3$, \, $\Delta N_{\rm eff}^{\rm BBN}\neq 0$ \\ 
        \hline \hline
    \end{tabular}
    \caption{
    Summary of the different scenarios considered as described in the text. 
    }
    \label{tab:cases}
\end{table}
%

\section{Cosmological Implications and Formalism}
\label{sec:formalism}

\emph{Cosmological Implications:} 
The cosmological implications of these light neutrino-philic bosons are governed by their decay rate into neutrinos. In particular, the ratio between the decay rate of $X$ into neutrinos and the Hubble parameter at $T\simeq m_X/3$ determines whether or not the $X$ boson thermalizes in the early Universe. In a radiation dominated Universe, this ratio can be parametrized by:
\begin{align}\label{eq:Keff_param}
    K_{\rm eff} &\equiv \left(\frac{\lambda_\nu}{4\times 10^{-12}}\right)^2\,\left(\frac{\rm keV}{m_X}\right) \\
    &\simeq \left.\frac{3\,\left<\Gamma (\bar{\nu}\nu \to X )\right>}{H}\right|_{T_\nu = m_X/3}  \, \nonumber ,
\end{align}
where $\left<\Gamma(\bar{\nu}\nu\to X)\right>$ is the thermally averaged inverse decay rate. 
For $K_{\rm eff} \gtrsim 1$ the $X$ boson thermalizes with the neutrinos in the early Universe via decays and inverse decays out of neutrinos\footnote{Processes such as $XX\leftrightarrow \bar{\nu}\nu$ are only effective for $\lambda_\nu \gtrsim 10^{-7}$ and as can be seen from Eq.~\eqref{eq:Keff_param} we will be interested in much smaller couplings.}. Thermalization has two important cosmological consequences:

\begin{enumerate}
    \item \textit{Non-standard expansion at $T_\nu \lesssim m_X$} -- If the $X$ boson thermalizes with neutrinos it will represent a non-negligible fraction of the energy density of the Universe. In particular, the $X$ boson will behave as radiation until $T_\nu \sim m_X$ but after it will start redshifting like matter and decay. This leads to a non-standard expansion history during this time, and to an enhanced value of $N_{\rm eff}$ at the time of recombination (provided that $m_X$ has decays before recombination). 
    
    \item \textit{Suppression of neutrino free streaming} -- The new interactions between neutrinos and the $X$ particle tend to homogenize the neutrino fluid, suppressing neutrino free streaming. This has important consequences for CMB observations as highlighted in the introduction.  
\end{enumerate}

\emph{Background Thermodynamics:} 
The exact description of the thermodynamic evolution of the Universe in the presence of a light boson interacting with neutrinos can be found by solving the Liouville equation for the distribution function of neutrinos and the $X$ boson. 
This is numerically very costly, but Ref.~\cite{EscuderoAbenza:2020cmq} explicitly demonstrated that for scenarios where the $X$ boson interacts efficiently with neutrinos, namely for $K_{\rm eff} \gtrsim 10^{-3}$, the thermodynamics can be accurately described by simple ordinary differential equations tracking the temperature and chemical potential of the neutrinos and the new light boson. 
These equations are explicitly outlined in Appendix~\ref{SM:theory}. 
\begin{figure*}[t]
\centering
\begin{tabular}{cc}
\hspace{-0.5cm} \includegraphics[width=0.47\textwidth]{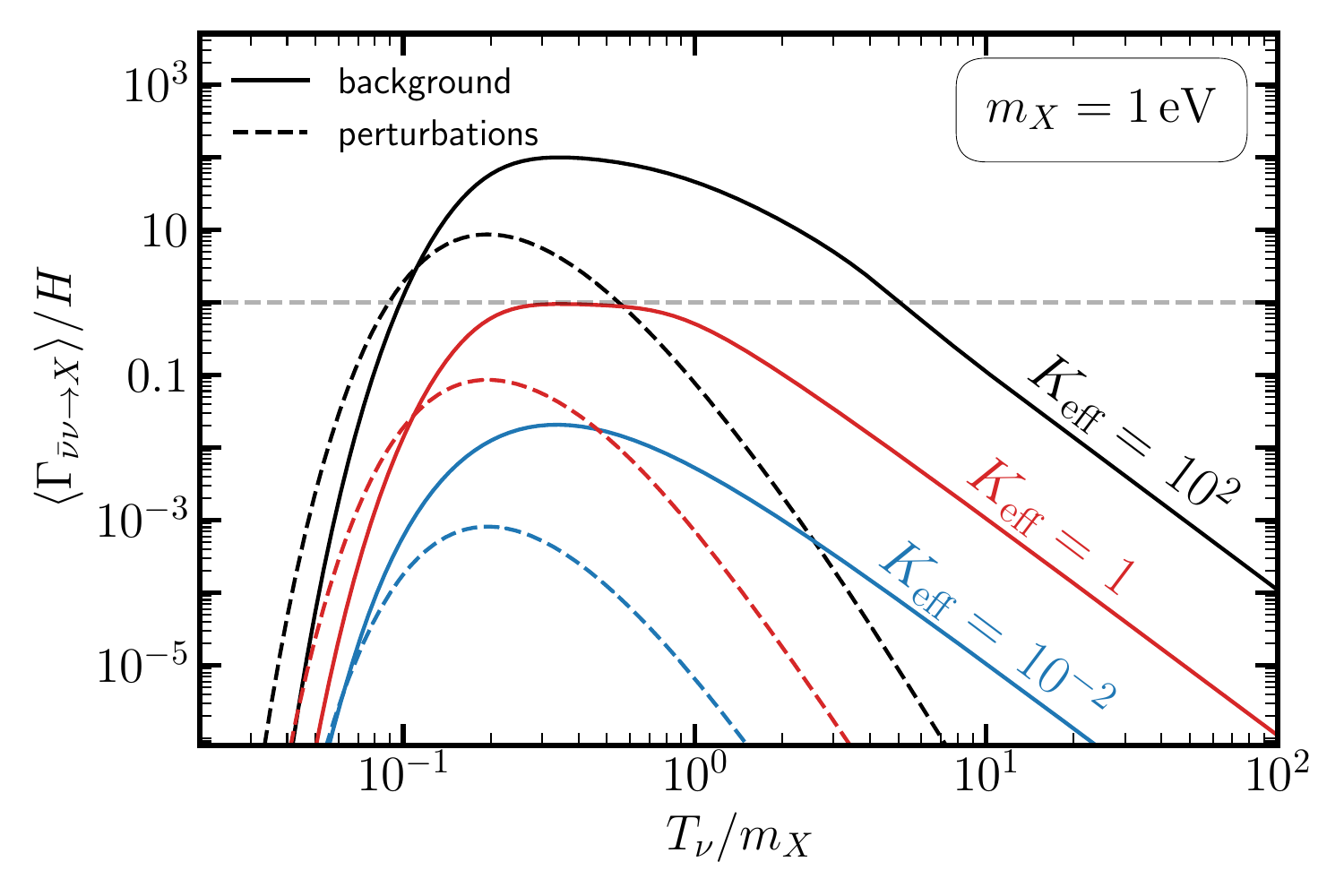} & 
\hspace{-0.5cm} \includegraphics[width=0.47\textwidth]{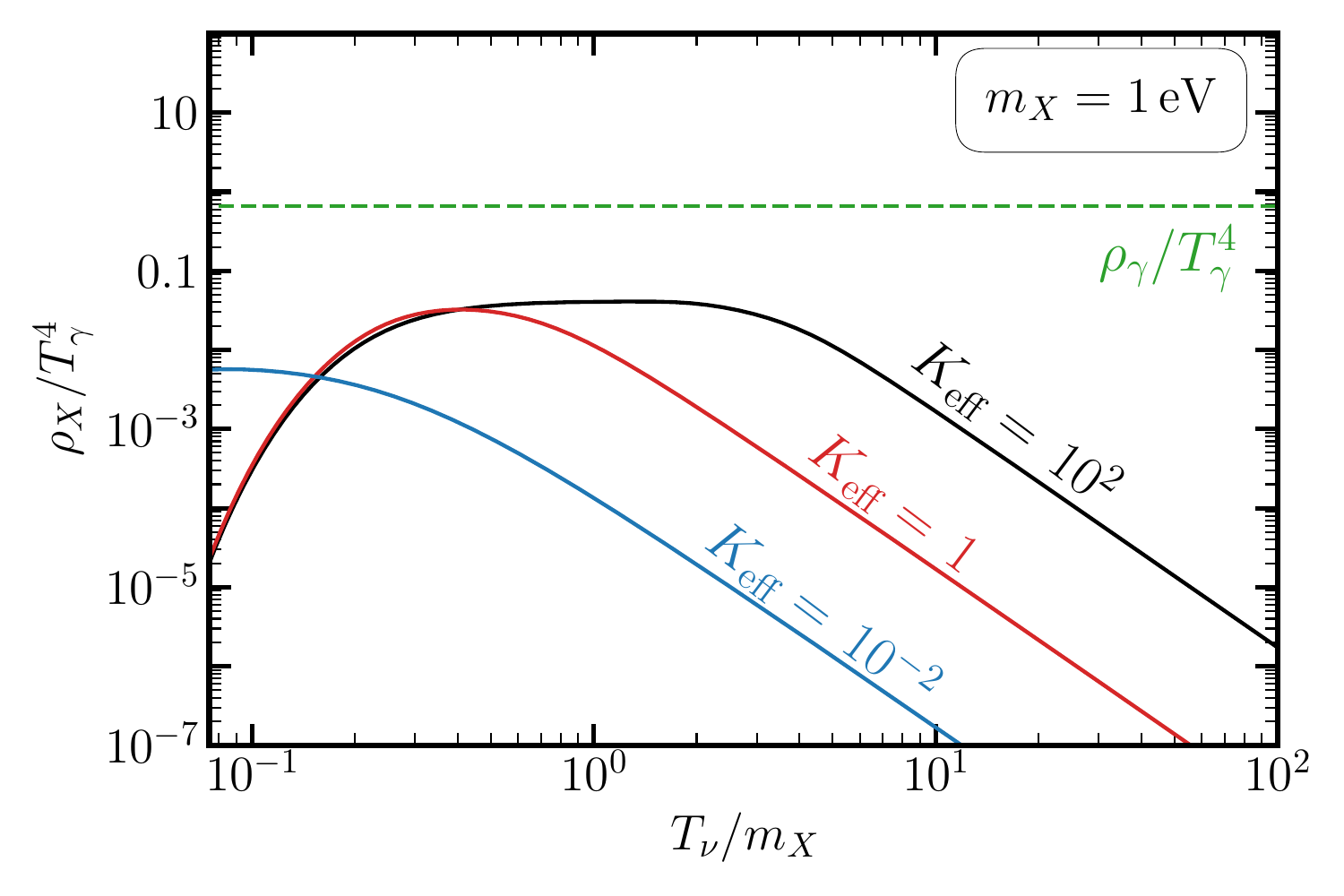}
\end{tabular}
\vspace{-0.4cm}
\caption{
\textit{Left:} Effective interaction rates at the background level (solid lines) as well as at the perturbation level  (dashed lines) for different values of $K_{\rm{eff}}$. The scenario considered consists of all $3$ neutrinos interacting with the scalar type boson.
\textit{Right:} Evolution of the normalized $X$ boson energy density for the same scenarios as before. For reference, we highlight in dashed the photon energy density.
}
\label{fig:Gamma_rhoevol}
\end{figure*}
In the left panel of Figure~\ref{fig:Gamma_rhoevol} we highlight the thermally averaged inverse decay rate ($ \left<\Gamma_{\bar{\nu}\nu\to X}\right> = \delta\rho_X/\delta t|_{\bar{\nu}\nu \to X} /\rho_\nu$) normalized to the Hubble parameter for a $m_X = 1\,{\rm eV}$ boson with $g_X = 1$. We show the evolution for several values of $K_{\rm eff} = 100\,,1\,,10^{-2}$ representing cases where thermal equilibrium is well established, where thermal equilibrium is only slightly reached, and where the $X$ boson does not thermalize, respectively. The energy density evolution for the $X$ particle for each of these cases is highlighted in the right panel of Figure~\ref{fig:Gamma_rhoevol}. From this figure we can clearly see that for $K_{\rm eff} \gtrsim 1$ the $X$ boson thermalizes with neutrinos and its thermodynamic evolution is dictated by thermal equilibrium. On the other hand, for $K_{\rm eff} < 1$ thermal equilibrium is not established which leads to out of equilibrium decays. The evolution at $T_\nu \lesssim m_X/3$ will lead in all cases to a non-standard expansion history. 

For $K_{\rm eff}\gg 1$ and for $m_X \gtrsim 10\,{\rm eV}$ thermal equilibrium dictates what is the value of the neutrino energy density after the $X$ particle has decayed away. By assuming thermal equilibrium and tracking the number and entropy densities of the neutrinos and $X$ species (see~\cite{EscuderoAbenza:2020cmq}), we can calculate the minimum values of $\Delta N_{\rm eff}$ at recombination for the scenarios (a)-(d). These results are outlined in Table~\ref{tab:DNeffvals}.
For $X$ being a scalar mediator one expects $\Delta N_{\rm eff}^{\rm CMB} = 0.08-0.12$ and for the vector mediator case $\Delta N_{\rm eff}^{\rm CMB} = 0.15-0.24$. 
\begin{table}[t]
    \centering
    \begin{tabular}{c|c} \hline \hline
        Model & $\,\,\Delta N_{\rm eff}^{\rm CMB} \,\,$  \\ \hline \hline
       $\,\,$ Case (a), $N_{\rm int} = 3$, $g_X = 1$  $\,\,$  & 0.12 \\  \hline 
       Case (b), $N_{\rm int} = 3$, $g_X = 3$ & 0.24 \\  \hline 
       Case (c), $N_{\rm int} = 1$, $g_X = 1$ & 0.08 \\  \hline 
       Case (d), $N_{\rm int} = 1$, $g_X = 3$ & 0.15 \\  \hline \hline
    \end{tabular}
    \caption{
    Minimum contributions to $\Delta N_{\rm eff}$ at the time of recombination resulting from the thermalization and subsequent decay of the $X$ neutrino-philic boson. This corresponds to $K_{\rm eff}\gg 1$ and $m_X \gtrsim 10\,{\rm eV}$.
    }
    \label{tab:DNeffvals}
\end{table}
We note that these values are similar to Planck's $1\sigma$ sensitivity to $N_{\rm eff}$, and thus an accurate treatment of this modified expansion history is needed to analyze the latest data.

\begin{figure}[t]
\centering
\includegraphics[width=0.47\textwidth]{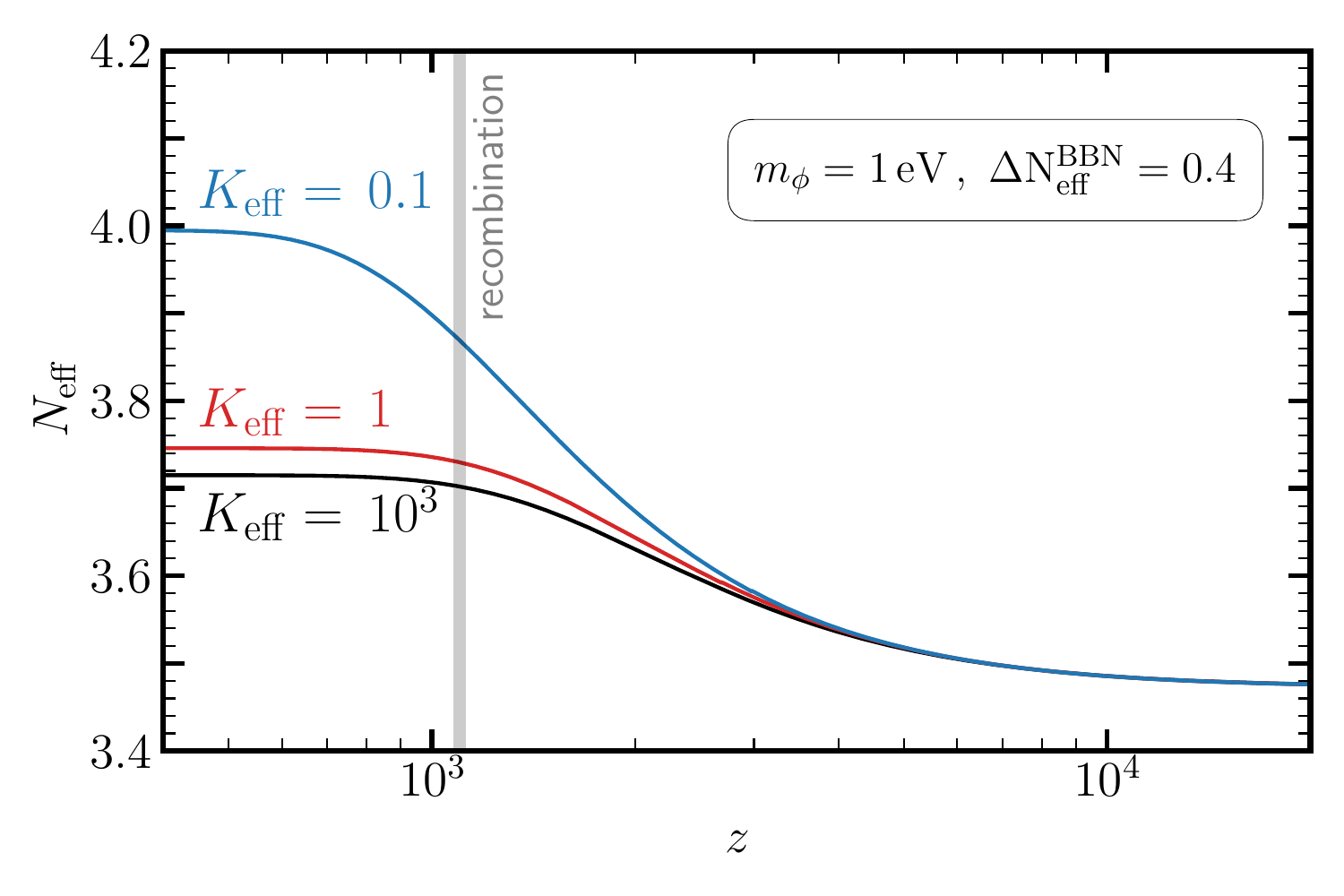} 
\vspace{-0.4cm}
\caption{
Evolution of $N_{\rm eff}$ for the case of a scalar interacting with three neutrinos with a primordial contribution to $\Delta N_{\rm eff}^{\rm BBN} = 0.4$. 
We notice that the value of $N_{\rm eff}$ always increases and that for small $K_{\rm eff}$ it increases significantly due to very out of equilibrium decays of the $X$ particle.
}
\label{fig:Neff_example}
\end{figure}

In the event that a primordial population of bosons already exists at the time of BBN, the process of thermalization at late times, i.e. near recombination, can significantly increase $\Delta N_{\rm eff}$.
For this reason, we differentiate the abundance of the new bosonic species at BBN and recombination using $\Delta N_{\rm eff}^{\rm BBN}$ and $\Delta N_{\rm eff}^{\rm CMB} $. 
We illustrate the evolution of $\Delta N_{\rm eff}^{\rm CMB} $ assuming a primordial abundance of $\Delta N_{\rm eff}^{\rm BBN} = 0.4$ in Figure~\ref{fig:Neff_example}. 
Two immediate conclusions can be drawn from this figure.
Firstly, the shift in $\Delta N_{\rm eff}$ between BBN and recombination can greatly exceed the values outlined in Table~\ref{tab:DNeffvals}. 
Secondly, $\Delta N_{\rm eff}$ increases dramatically for $K_{\rm eff}\lesssim 1$.
This is because the $X$ boson becomes non-relativistic and its delayed decay leads to a significant increase of the relative energy stored in this species.
Consequently, scenarios with $\lambda_\nu \to 0$ and $\Delta N_{\rm eff}^{\rm BBN} \neq 0$ lead to a drastically distinct phenomenology compared to $\Lambda$CDM.
Although, the effect of neutrino-free streaming suppression is negligible, these scenarios will be tightly constrained from the increase in $\Delta N_{\rm eff}$.
\emph{Cosmological Perturbations:}
In order to track the cosmological perturbations of the fluids describing neutrinos and the neutrino-philic boson $X$, we rely on several approximations. 
First, we treat the two interacting fluids as coupled, as done in past literature~\cite{Barenboim:2020vrr,Chen:2022idm}.
This implies that we can evolve the perturbations jointly. 
In the limit that the interactions are sufficiently strong this approximation is by definition valid.
On the other hand, in the weak interaction limit, we also expect the approximation to be valid, because the perturbation equations in this case are equivalent to two decoupled fluids.

The second approximation adopted here enters in the collision term describing the $1\leftrightarrow 2$ interactions between the neutrinos and the $X$ boson.
Following Ref.~\cite{Chen:2022idm} we assume:
(1) Maxwell-Boltzmann statistics, 
(2) that the background momentum dependence of the neutrino distribution is not strongly time dependent, and 
(3) that the perturbation generated by gravity is universal to all the species involved.
We expect all these approximations to hold  in our scenario.

Finally, we treat neutrinos as being massless. 
This assumption significantly simplifies the evolution of the neutrino perturbations. Since current Planck data is consistent with massless neutrinos, setting an upper limit on the sum of neutrino masses at the level of $\sum m_\nu < 0.12\,{\rm eV}$~\cite{planck}, we believe this approximation does not significantly alter our results. Nevertheless, a more thorough treatment including neutrino masses would be of interest, and thus we leave this for future work.

Under the approximations listed above, the equations describing the joint neutrino$+$boson system in synchronous gauge read~\cite{Ma:1995ey}:
\begin{subequations}\label{eq:Hierarchy}
\begin{align}
\label{eq:pert_delta}
\dot{\delta} &= - (1+w) \left(\theta+{\dot{h}\over 2}\right) - \mathcal{H} \left(c_s^2 - w \right)\delta  \,,\\
\label{eq:pert_theta}
\dot{\theta} &= - \mathcal{H} (1-3w)\theta - {\dot{w}\over 1+w}\theta + {c_s^2 \over 1+w}\,k^2\delta - k^2 \sigma\,,\\
\label{eq:F2}
\dot{F} {}_{2} &= 2\dot{\sigma} = \frac{8}{15}\theta-\frac{3}{5}kF_{3}  +\frac{4}{15}\dot{h}+\frac{8}{5}\dot{\eta}  - 2\, a\, \Gamma_{\rm NF\,2} \, {\sigma}   ,  \\
\label{eq:Fl}
\dot{F}_{\ell} &= \frac{k}{2\ell + 1} \left[ \ell \, {F}_{\ell-1} - (\ell +1){F}_{\ell+1}   \right] - a \,  \Gamma_{\rm NF\,\ell} \, {F}_{\ell}  \,,\,  {\rm for} \,\,\ell \geq 3 \, .
\end{align}
\end{subequations}
Here, derivatives are taken with respect to conformal time, $\mathcal{H}$ is the conformal Hubble parameter, $h$ and $\eta$ represent the metric perturbations, $a$ is the scale factor, $\omega = p/\rho$ is the equation of state of the system, $c_s^2 = dp/d\rho$ is the sound speed squared, $k$ defines the given Fourier mode, $\delta$ and $\theta$ are the energy and velocity perturbations respectively,  $F_{\rm \ell}$ represents the $\ell$ moment of the perturbed distribution function, and the neutrino free streaming suppression rate is given by is~\cite{Chen:2022idm}: 
\begin{align}
\label{eq:GNF}
    \Gamma_{\rm NF\,{\ell}} &= - \alpha_\ell \,  \frac{g_X}{4\pi^2} \frac{m_X T_\nu^3}{\rho_X+\rho_\nu}  \Gamma(X\to {\bar \nu}\nu) \,\left( \frac{m_X}{T}\right)^4  \,\mathscr{F}\left(\frac{m_X}{T_\nu}\right) \,.
\end{align}
In this expression we neglect the chemical potentials, which we explicitly checked to have negligible impact on observables.
The coefficients are given by~\cite{Chen:2022idm}
\begin{align}
    \alpha_\ell &\equiv (3\ell^4+2\ell^3-11\ell^2+6\ell)/32 \,,\\
    \mathscr{F}(x) &\equiv  \frac{1}{2} {\rm e}^{-x} \left( -1 + x - {\rm e}^{x}(x^2-2)\Gamma(0,x)\right)\,,
\end{align}
where $\Gamma(0,x)$ is the incomplete gamma function. At high temperatures $\Gamma_{\rm NF} \sim (m_X/T_\nu)^5\,\Gamma(X\to {\bar{\nu}}\nu) $ and at very small temperatures $\Gamma_{\rm NF} \sim e^{-m_X/T_\nu} \Gamma(X\to {\bar{\nu}}\nu)$. 
This neutrino free streaming rate is shown as a function of temperature in dashed lines in the the left panel of Figure~\ref{fig:Gamma_rhoevol}. 
We can clearly see that at high temperatures the scaling of $\Gamma_{\rm NF}$ is different to the background evolution.
Moreover at $T_\nu \sim m_X/3$, where the rate is maximal, it is a factor of $\sim 1/10$ smaller than the background equivalent.
It is actually easy to see that for $\Gamma_{\rm NF}/H > 1$, $F_\ell \to 0$ exponentially fast, which strongly reduces neutrino free streaming. 
\begin{figure*}[t]
\centering
\begin{tabular}{cc}
\hspace{-0.5cm} \includegraphics[width=0.47\textwidth]{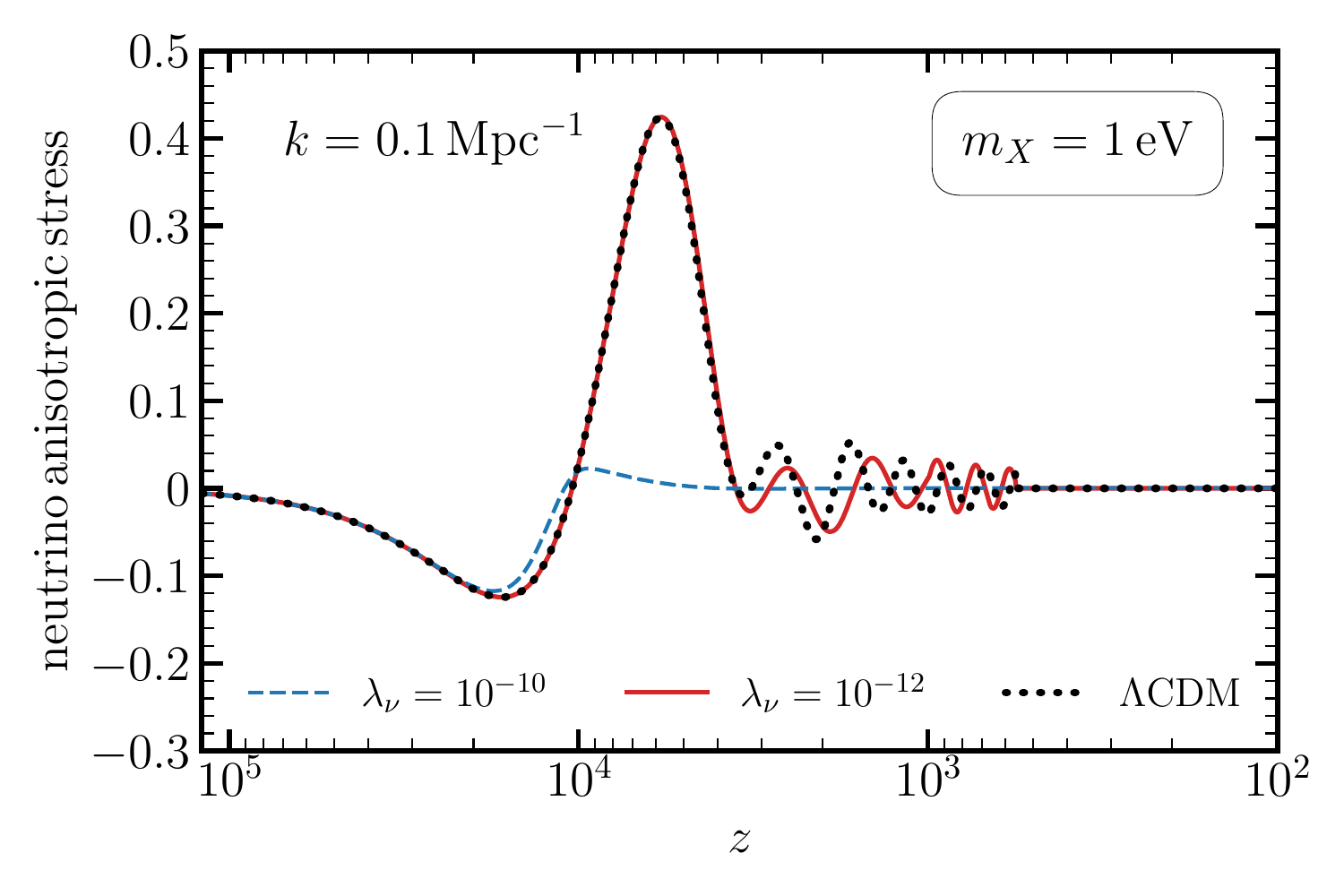}    &
\hspace{-0.5cm} \includegraphics[width=0.47\textwidth]{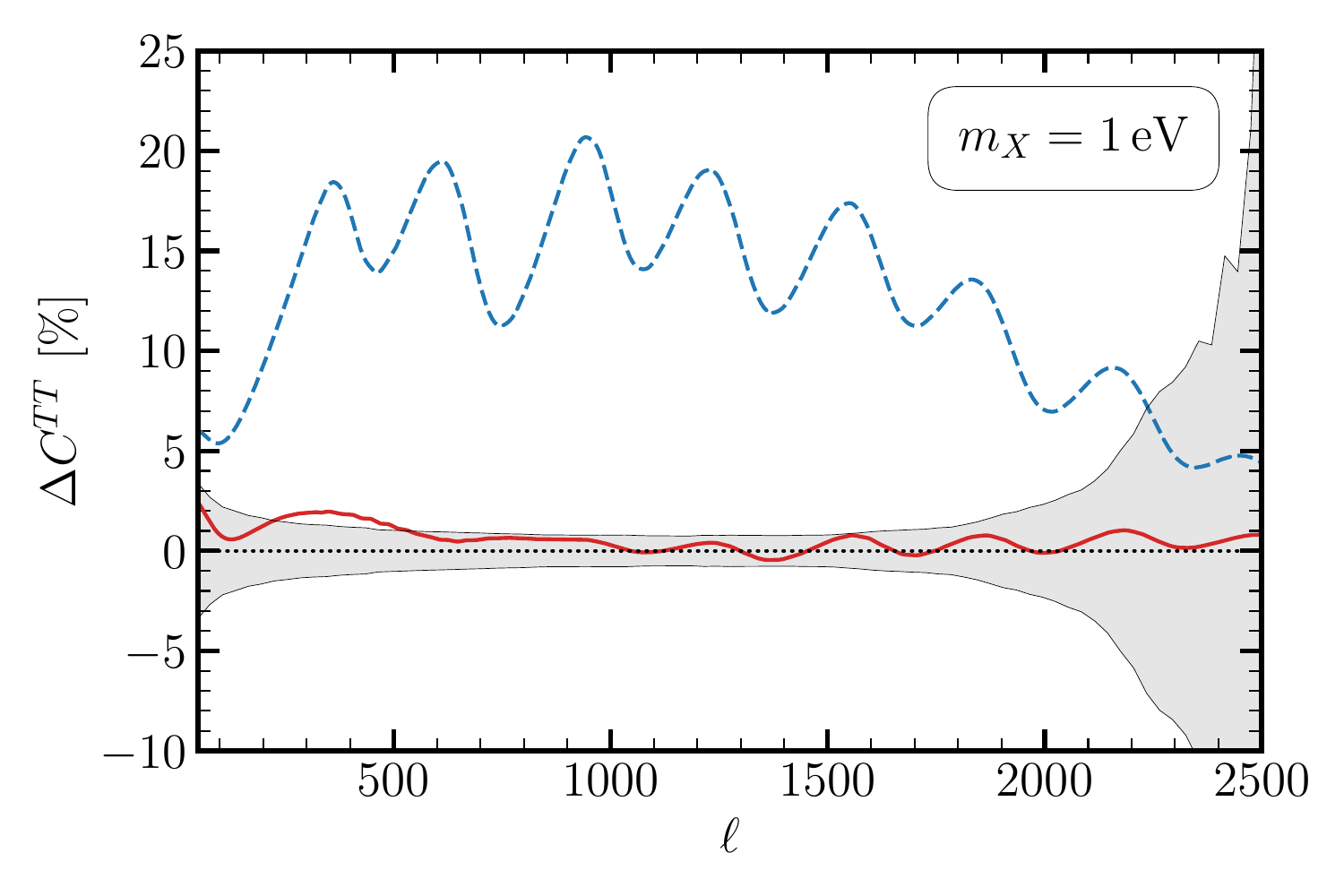}
\end{tabular}
\vspace{-0.4cm}
\caption{
\textit{Left panel:} Evolution of the neutrino anisotropic stress for a mode of $k = 0.1\,{\rm Mpc}^{-1}$ for $\Lambda$CDM and an scenario with $N_{\rm int} = 3$ neutrinos interacting with a scalar with different coupling strengths. \textit{Right panel:} Relative difference of the TT power spectrum in a majoron cosmology with respect to $\Lambda$CDM as a function of multipole $\ell$. We show for reference the size of the Planck error bars. The comparison has been made with fixed standard cosmological parameters. We can clearly appreciate how the strong damping of the neutrino anisitropic stress on the left hand side is strongly related with a strong change on the power spectra.   
}\label{fig:CLs_shear}
\end{figure*}
\begin{figure}[t]
\centering
\hspace{-0.5cm}\includegraphics[width=0.47\textwidth]{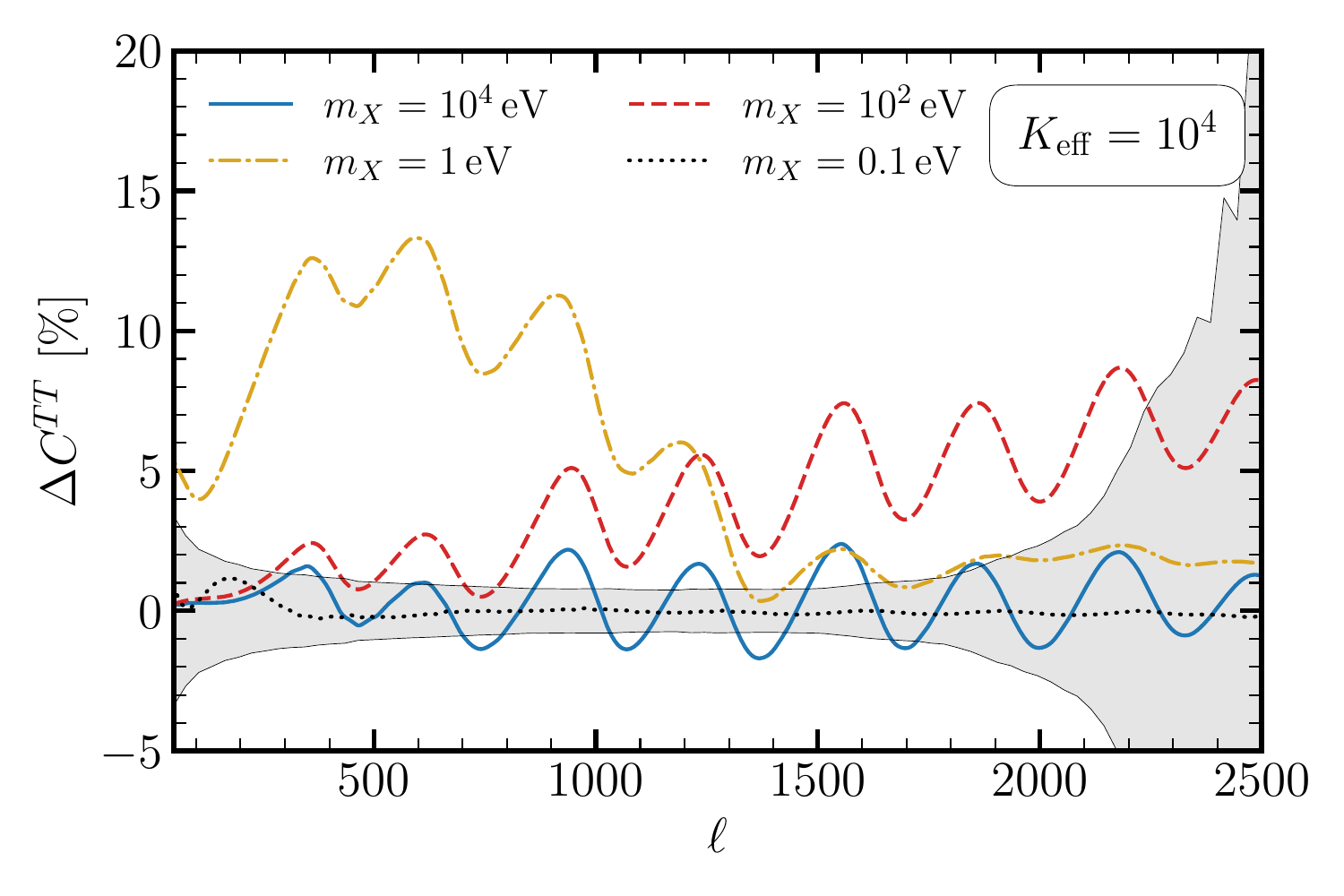} 
\vspace{-0.4cm}
\caption{
Fractional difference on the TT power spectrum with respect to $\Lambda$CDM for the case of a scalar particle interacting efficiently with neutrinos, $K_{\rm eff} = 10^4$, see Eq.~\eqref{eq:Keff_param}. We show the results for different values of $m_X$. 
}\label{fig:CLs_masschange}
\end{figure}
\emph{Numerical Implementation in CLASS:} 
We track the impact of the neutrino-$X$ interactions on the CMB power spectrum by modifying the cosmological Boltzmann code CLASS~\cite{Lesgourgues:2011re,Blas:2011rf}. 
The code is available on \gitlink.
It can also help to study the thermodynamic evolution of different BSM scenarios. 

In the left panel of Figure~\ref{fig:CLs_shear} we show the evolution of the neutrino anisotropic stress associated with a mode of $k = 0.1\,{\rm Mpc}^{-1}$ as a function of redshift. 
We choose $k = 0.1\,{\rm Mpc}^{-1}$ because it is the largest wave number well probed by CMB observations.
The evolution for different, smaller wave numbers are shown in Figure~\ref{fig:SM_delta_shear} of the appendix.
From Figure~\ref{fig:CLs_shear} we can clearly see how the decays and inverse decays of $X$ reduce the neutrino anisotropic stress. 
In the right panel of the same figure we also show the relative impact on the temperature power spectrum $C_\ell^{TT}$ compared to $\Lambda$CDM.
The impact on the observable $C_\ell^{TT}$ spectrum can go well above the level of the $1\sigma$ relative error bars, as indicated by the grey band.

In Figure~\ref{fig:CLs_masschange} we show the CMB temperature power spectrum for different values of $m_X$, taking $N_{\rm int} = 3$, $g_X=1$, and fixing $K_{\rm eff} = 10^4$. This corresponds to a scenario where the $X$ particle interacts very efficiently with neutrinos, and thermal equilibrium is reached  at $T\sim 30\times m_X$. From this plot we can appreciate a number of interesting features: firstly, we notice that for $m_X \lesssim 0.1 {\rm eV}$ the impact on the CMB power spectrum is not significant. This is because the non-standard expansion history occurs after recombination, and owing to the high temperature suppression in the collision term, neutrino free streaming is not significantly altered before recombination. We notice that the most significant effect is for bosons with $1\,{\rm eV} \lesssim m_X \lesssim 100\,{\rm eV}$. 
This is because the interaction rate of these bosons is maximal during the window of redshift to which the CMB is sensitive, i.e. $2000\lesssim z\lesssim 10^5$. 
Finally, for the case with heavy mediator, $m_X = 10\,{\rm keV}$, the boson can not alter late-time free streaming, since it will have decayed already at higher redshift.
This means that the observed effect purely corresponds to a shift in $N_{\rm eff}$ of $0.12$ (see Table~\ref{tab:DNeffvals}).

\section{CMB Data Analysis and Results}\label{sec:results}

\emph{Cosmological Data and Analysis:} 
We perform MCMC analyses with {\tt MontePython}~\cite{Audren:2012wb,Brinckmann:2018cvx} on each of the models listed in Table~\ref{tab:cases}.
For the likelihood we use data from {\tt Planck2018+BAO} data~\cite{planck,Aghanim:2019ame}.
In particular, this includes the temperature and polarization power spectra, as well as the lensing likelihood, from Planck~\cite{Aghanim:2019ame}, and the 6DF galaxy survey~\cite{Beutler:2011hx}, the MGS galaxy sample of SDSS~\cite{Ross:2014qpa}, and the CMASS and LOWZ galaxy samples of BOSS DR12~\cite{Alam:2016hwk,Vargas-Magana:2016imr,Ross:2016gvb,Beutler:2016ixs}.  
In order to investigate the extent to which these scenarios could explain or ameliorate the Hubble tension we perform additional MCMC analyses including a Gaussian likelihood on  $H_0 = 73.30\pm 1.04\,{\rm km/s/Mpc}$~\cite{Riess:2021jrx}. These results are used to replicate the three statistical criteria (described in detail below) introduced in the `$H_0$ Olympics'~\cite{Schoneberg:2021qvd}.
This comparison allows to establish the relative success and failure of the models of Table~\ref{tab:cases} in relation to other proposed solutions.

For the standard cosmological parameters and the nuisance parameters of the Planck likelihood we use the same priors as the Planck collaboration. For the mass and coupling of the neutrino-philic bosons we adopt log priors over the range:
\begin{align}
    \log_{10} (\lambda_\nu)  &\in [-15,-6] \\
    \log_{10} (m_X/{\rm eV}) &\in  [-1.0,3.5]  \, .
\end{align}
The lower bound on $m_X$  corresponds to twice the minimum mass of the heaviest neutrino, $2\sqrt{|\Delta m_{\rm atm}^2|} \simeq 0.1\,\rm{eV}$. 
For the case of the $X$ boson interacting with $N_{\rm{int}} < 3$ neutrino families, the prior range is extended to $\log_{10} (m_X/{\rm eV}) \in  [-4,3.5]$ as one of the neutrinos could be much lighter and thus open up parameter space for lighter $X$ bosons.
The lower limit in this case is chosen to be sufficiently small such that the interaction rate is never effective to thermalize the $X$ boson. 
We also introduce a specific upper limit on $\lambda_\nu = 10^{-6}$.
This is because at larger couplings two-to-two processes ($X X \leftrightarrow \nu\bar{\nu}$), which are not captured by our treatment, begin to become relevant. 
On the other hand, the lower limit in the coupling is chosen to be sufficiently small that the $X$ boson is effectively fully decoupled from the neutrino sector.
In this limit, $\lambda_\nu \to 0$, $\Lambda$CDM is recovered.
At sufficiently large masses, the $X$ boson decays at high redshift, producing a shift in $\Delta N_{\rm eff}$ without altering neutrino free streaming -- our upper bound on the mass is set by the fact that this effect is the same for $m_X \gtrsim 1\,\rm{keV}$ (assuming a sufficiently large coupling such that the bosons thermalize). 
Finally, in some of the scenarios we also allow for a non-zero initial abundance of the $X$ particle.
We parameterize it by $\Delta N_{\rm eff}^{\rm{BBN}}$ and adopt a flat, linear prior over the range 
\begin{align}
    \Delta N_{\rm eff}^{\rm{BBN}}\in [0,0.7]\,.
\end{align}

Performing the MCMC analysis with the likelihoods and priors as described above leads to the result of Figure~\ref{fig:bounds_majoron} which combines cases (a)-(d) of Table~\ref{tab:cases}.
These runs contain a total of $N\sim 2\times 10^6$ samples. 
The $3\sigma$ exclusion region is obtained by binning the points in $\log_{10}(m_X/\rm{eV})$, and in each bin determining the coupling $\lambda_\nu$ for which $99.7\%$ of the samples have $\lambda_\nu \leq \lambda_{\rm limit}$.
A particularly interesting result is obtained for the scenario (c), i.e. the scalar boson $X$ which interacts with $N_{\rm int} = 1$ neutrino family.
In this scenario, we find a slight statistical preference for non-zero neutrino interactions; we note, however, that the $\Lambda$CDM limit is also favored at the  $1\sigma$ level, implying the statistical preference for this best-fit region is not remarkably significant. This region can be seen more clearly in Figure~\ref{fig:SM_PointsParameterspace}, where the MonteCarlo samples are explicitly shown. This best fit region of parameter space roughly corresponds to:
\begin{align}
\label{eq:GNF_1sigma}
    \Gamma_{\rm NF}/H(z) = 1 \,\,\text{at} \,z = 1100-3500\,,
\end{align}
namely, this preferred region of parameter space corresponds to scenarios where the neutrino anisotropic stress starts to be damped right before recombination, $1100 \lesssim z \lesssim 3500$. This is highlighted by the red region labelled `best fit region' in Figure~\ref{fig:bounds_majoron}.

We note that we do not find such a preferred region of parameter space for scenario (d) with a gauge boson interacting with a single neutrino species. The suppression of neutrino free-streaming is very similar to the case of a scalar, and thus we attribute the lack of preference for parameter space to the fact that the vector boson leads to a substantially enhanced expansion history for which Planck is sensitive to, see the lower row of Table~\ref{tab:DNeffvals}. 

\begin{center}
\textbf{Implications for the Hubble Tension}
\end{center}

It has been shown in~\cite{Escudero:2019gvw,EscuderoAbenza:2020egd,Escudero:2021rfi} that models with neutrino $X$-boson interactions can have the potential to significantly ameliorate the Hubble tension for two main reasons: 1) the $X$-neutrino interactions can lead to a non-trivial enhancement of the expansion history near recombination, 2) there exists a level of degeneracy between the impact of the damping of neutrino free streaming and an enhanced value of $N_{\rm eff}$ which allows for additional radiation without spoiling the fit to the data from Planck. 
In particular, the detailed statistical analysis of the `$H_0$ Olympics'~\cite{Schoneberg:2021qvd} awarded the model with a silver medal.
However, as mentioned above, the original implementation of this model relied on numerous approximations. For this reason, we revisit the three `$H_0$ Olympics' criteria using the improved analysis developed here. These criteria include:
\begin{enumerate}
    \item The Gaussian Tension, given by
    \begin{equation}
\frac{\overline{H_0}_\mathcal{C} - \overline{H_0}_{\rm SH_0ES}}{\sqrt{\sigma_\mathcal{C}^2 + \sigma_{\rm SH_0ES}^2}}  \,,
    \end{equation}
    where $\overline{H_0}_i$ and $\sigma_i$ are the central value and the uncertainty on the inferred value $H_0$.
    The index $i=\{\mathcal{C}, {\rm SH_0ES} \}$ refers to the cosmologically inferred value (using Planck and BAO) or the value measured by $\rm{SH_0ES}$, $H_0 = 73.3 \pm 1.04\,{\rm km/s/Mpc}$.  
    \item The $Q_{\rm DMAP}$ (difference of the maximum a posteriori), given by
    \begin{equation}
        \sqrt{\chi^2_{{\rm min}, \mathcal{C}+{\rm SH_0ES}} - \chi^2_{{\rm min}, \mathcal{C}}} \, ,
    \end{equation}
    where the minimum $\chi^2$ is evaluated using a likelihood that does ($\mathcal{C}+\rm{SH_0ES}$) and does not contain ($\mathcal{C}$) the $\rm{SH_0ES}$ likelihood.
    \item Akaike Information Criterium (AIC), given by
    \begin{equation}
        \Delta {\rm AIC} = \chi^2_{\rm min, \mathcal{M}} - \chi^2_{\rm min, \Lambda{\rm CDM}} + 2 (N_{\mathcal{M}} - N_{\Lambda CDM}) \, ,
    \end{equation}
    where $\mathcal{M}$ refers to the model under consideration and $N$ corresponds to the number of free parameters of that model. Here, the $\chi^2_{\rm min}$ values are obtained using a likelihood that includes the Gaussian contribution from $\rm{SH_0ES}$. 
\end{enumerate}
Each criteria is intended to address a slightly different question -- we refer the interested reader to~\cite{Schoneberg:2021qvd} for a broader overview of the benefits and drawbacks of each. 
The results of each model are summarized in Table~\ref{tab:H0results}.
There we also show for comparison the $\Lambda$CDM result and the simple scenario containing free streaming dark radiation as parameterized by $\Delta N_{\rm eff}$. 
Interestingly, none of the models investigated show a significant reduction in the cosmological tension, with the most successful of them only reducing it to the $3.2\sigma$ level (in comparison with $4.5\sigma$ for $\Lambda$CDM). 
This result obtained here represents a degradation compared to what was found in previous works~\cite{Escudero:2019gvw,EscuderoAbenza:2020egd,Escudero:2021rfi}. The main reason for this deviation is due to the refined collision term included here, see Eq.~\eqref{eq:GNF}, which reduces the damping  of neutrino free streaming with respect to the approximation of ~\cite{Escudero:2019gvw,EscuderoAbenza:2020egd,Escudero:2021rfi} at $T \gg m_X$.
In particular, the full collision term helps to break the partial degeneracy between the damping of the neutrino free streaming at high redshift and the enhancement of $\Delta N_{\rm eff}$.  
\begin{table}[t]
    \centering
    \begin{tabular}{c|c|c|c} \hline \hline
        Model/Metric & Gaussian Tension & $Q_{\rm DMAP}$ & $\Delta$AIK  \\ \hline \hline
        $N_{\rm int} = 3$, scalar & 3.71  & 3.20 & 0.67 \\  \hline 
        $N_{\rm int} = 1$, scalar & 3.73 & 4.10 & 2.22 \\  \hline 
        $N_{\rm int} = 3$, vector & 3.72  & 3.71 & 2.44 \\  \hline 
        Dark Radiation  & 3.76  & 3.96 & -1.0 \\   \hline  $\Lambda$CDM  & 4.55  & 4.56 & 0 \\  \hline  \hline
    \end{tabular}
    \caption{Comparison of tension metrics for three different models, a simple model with free streaming dark radiation and $\Lambda$CDM. Note that the tension is slightly below $5\sigma$ in $\Lambda$CDM because we are considering purely massless neutrinos for simplicity.}
    \label{tab:H0results}
\end{table}
%
%

\section{Additional Constraints}
\label{sec:otherconstraints}

The models we have discussed in the main text are subject to additional constraints coming from other cosmological probes, emission from astrophysical objects, and laboratory searches. In this section we briefly highlight the origin of each constraint shown in Figure~\ref{fig:bounds_majoron}.

\textit{Laboratory Constraints:}
In the two benchmark particle physics models we consider, see Eqns.~\eqref{eq:Lag_scalar}-\eqref{eq:Lag_vector}, the coupling of the new boson to neutrinos is constrained by a different set of laboratory constraints. In the case of $X$ being identified as a light scalar, its coupling to neutrinos can give rise to double beta decay along the emission of a scalar. The latest constraints on $\lambda_\nu$ from the non-observation of such a process from the EXO-200 experiment reads: $\lambda_\nu < 0.9\times 10^{-5}$~\cite{Kharusi:2021jez}. In the case of $X$ being a light $U(1)_{L_\mu-L_\tau}$ gauge boson, we adopt a nominal value of kinetic mixing induced at 1-loop by muons and taus, $\epsilon \simeq -g_{\mu-\tau}/70$~\cite{Pospelov:2008zw}. 
The presence of this mixing can in turn change the scattering rate of neutrinos and electrons, which has been precisely measured by Borexino~\cite{Bellini:2011rx}. 
For $m_X \lesssim {\rm MeV} $, the coupling is constrained to be $g_{\mu-\tau} < 4\times 10^{-5}$~\cite{Gninenko:2020xys,Kamada:2015era,Harnik:2012ni}. Both the EXO-200 and Borexino bounds are shown in Figure~\ref{fig:bounds_majoron}.

\textit{Supernova Bounds:} 
Despite being very weakly coupled, the neutrino-philic bosons considered in this work can be copiously produced in extreme astrophysical environments such as supernovae.
If so, these particles can modify the energy and temporal distributions of the neutrino flux arriving on Earth.
In particular, in the majoron model the neutrino coalescence $\bar{\nu}\nu \to \phi$ can produce a delayed high-energy neutrino signal~\cite{Fiorillo:2022cdq,Heurtier:2016otg,Brune:2018sab,Akita:2022etk}.
The non-observation of such a signature in the measured neutrino flux from $\rm{SN1987A}$~\cite{Kamiokande-II:1987idp,Bionta:1987qt,Alekseev:1988gp} leads to the following constraint~\cite{Fiorillo:2022cdq}:
\begin{align}
5\times 10^{-10} <    \lambda_\nu \frac{m_X}{\rm MeV} \sqrt{g_X} < 1.3\times 10^{-7}  \,,
\end{align}
for $10\,{\rm  keV} \lesssim m_X \lesssim 1\,{\rm MeV}$.

On the other hand, the high densities present at supernovae induce flavour and helicity dependent effective neutrino masses. Therefore, for masses $m_X\lesssim 10 \, {\rm keV}$, the process ${\bar \nu} \to {\nu} X $ in kinematically allowed~\cite{Kachelriess:2000qc, Farzan:2002wx}.
Including these processes one finds constraints at the level of
\begin{align}
5\times 10^{-7}  \lesssim    \lambda_\nu  \lesssim 3\times 10^{-5} \, .
\end{align}

The SN1987A bound for a $U(1)_{L_\mu-L_\tau}$ gauge boson were derived in~\cite{Escudero:2019gzq,Croon:2020lrf}. The emission of gauge bosons of $m_{Z'}<{\rm MeV}$ is dominated by semi-Compton processes $\mu\gamma\to \mu Z'$ and the constraint imposed by the observation of the SN1987A signal is at the level of $g_{\mu-\tau} \lesssim 10^{-9}$~\cite{Croon:2020lrf}.

\textit{Star Cooling:} 
A light $U(1)_{L_\mu-L_\tau}$ gauge boson with the canonical kinetic mixing  interacts with charged matter, and thus can be produced in stars. Should these particles be produced, they can free stream out of the star, carrying away a sizeable amount of energy. 
Consequently, strong constraints can be derived by requiring that the stellar cooling rate is not significantly altered. 
Recasting the limits derived in~\cite{Hardy:2016kme} (see also~\cite{An:2013yfc} and~\cite{Li:2023vpv}) using the nominal kinetic mixing $\epsilon = - g_{\mu-\tau}/ 70$ yields the bound in Figure~\ref{fig:bounds_majoron}, labelled `Stars'.

\textit{BBN Bounds:} 
The production of new relativistic particles prior to BBN will enhance the value of $\Delta N_{\rm eff}$.
This modifies the expansion rate and in turn the prediction of the primordial element abundances. 
Current observations of the primordial abundances are consistent with $\Delta N_{\rm eff} \sim 0$. In particular, $\Delta N_{\rm eff}^{\rm{BBN}} \leq 0.41$ at $2\sigma$~\cite{Yeh:2022heq, Pisanti:2020efz}, and thus large deviations from this can yield strong constraints on the interactions with new particles. 

Limits were recently derived on the majoron by identifying the couplings for which  $\bar{\nu}\nu\to \phi$ lead to a shift in $\Delta N_{\rm eff}$ at the level of 0.5~\cite{Escudero:2019gvw}. Comparable constraints were derived on the $\mu-\tau$ gauge boson from the production of a primordial population via $\mu^+\mu^- \to Z'\gamma$ processes~\cite{Escudero:2019gzq}. These constraints are shown in Figure~\ref{fig:bounds_majoron} with the label `BBN'. 

\textit{CMB bounds on out of equilibrium decays:} 
The thermodynamic treatment of the neutrino-philic bosons used in this study is only capable of accounting for moderate departures of thermal equilibrium, namely for $K_{\rm eff} \gtrsim 10^{-3}$~\cite{EscuderoAbenza:2020cmq}. In the absence of a primordial abundance, the region of parameter space with $K_{\rm eff}\lesssim 10^{-3}$ is irrelevant as $K_{\rm eff}$ controls the production of $X$ particles and for such small $K_{\rm eff}$ the energy density of $X$ particles is negligible. However, even a small primordial abundance in the weakly coupled limit can yield strong observable consequences. 
The reason is that the primordial species can become non-relativistic prior to matter-radiation equality, dramatically increasing the relative energy density stored in this species before it undergoes an out-of-equilibrium decay into neutrinos. 
The detailed treatment of this scenario is rather intricate (see e.g.~\cite{Holm:2022eqq,Blinov:2020uvz}), and a full parameter space exploration is still lacking. 
In order to illustrate where these constraints would lie, we assume a primordial abundance at BBN of $\Delta N_{\rm eff}|_{\rm BBN} = g_X \times 0.027$ (corresponding to the minimal value predicted for a boson that was in thermal equilibrium at temperatures above the electroweak phase transition) and derive an approximate constraint by requiring that $N_{\rm eff} < 4$ at recombination. We did this by tracking the evolution of the $X$ boson energy density allowing for out of equilibrium decays and neglecting inverse decays (which are highly inefficient in this region of parameter space). In Figure~\ref{fig:bounds_majoron} this constraint is indicated by the pink region labelled `out of equilibrium decay' (and would exclude couplings {\emph{below}} this line).

\section{Summary, Conclusions and Outlook}\label{sec:conclusions}

In this work, we have presented an improved treatment of the cosmological evolution of weakly coupled neutrino-philic bosons with masses in the $\mathcal{O}({\rm eV})$ range. This work represents a significant improvement upon previously analyses~\cite{Escudero:2019gvw, Escudero:2021rfi}, which focused exclusively on the singlet majoron model and relied on a number of simplified approximations. 
Specifically, in this manuscript we present three updates:
\begin{enumerate}
    \item We have incorporated the thermodynamic evolution tracing the out-of-equilibrium thermalization of the neutrino-philic bosons directly in the Boltzmann solver CLASS. This allows for a more accurate and careful treatment of the neutrino-boson interactions across a wide array of parameter space.
    The developed code is made public on \gitlink.
    \item We have incorporated a recently derived collision term~\cite{Barenboim:2020vrr,Chen:2022idm}, which captures the impact of these interactions on the damping of the neutrino anisotropic stress.
    \item We generalize this analysis to include: interactions with one, two, or three neutrino species,  and both vector and scalar bosons. Our fiducial limits are  recasted in the terms of the singlet majoron model and the $U(1)_{L_\mu-L_\tau}$ gauge boson, but these limits can be easily interpreted in the context of many other neutrino-philic boson models.
\end{enumerate}

As shown in Figure~\ref{fig:bounds_majoron}, the limits derived using a combination of CMB and BAO data provide the strongest constraints to date across a range of masses near the $\mathcal{O}({\rm eV})$ scale. 
We have also revisited the extent to which neutrino-philic bosons can resolve the Hubble tension.
We show that the improved collision term, which is strongly suppressed in comparison to the previous approximations at $T \gg m_X$, significantly degrades the extent to which neutrino-philic bosons can ameliorate the tension. 

In the case of the majoron singlet model, there exists a slight preference in the data for non-zero majoron-neutrino interactions (at the $\sim 1\sigma$ level). This region of parameter space is expected to be fully probed in the near future by LiteBIRD~\cite{LiteBIRD:2020khw} thanks to a cosmic variance limited measurement of the large scale EE polarization power spectrum. Upcoming observations from the Simons Observatory~\cite{SimonsObservatory:2018koc} are expected to measure $N_{\rm eff}$ with a $1\sigma$ precision of 0.05. This will be an improvement by a factor of 4 as compared with Planck and will significantly improve sensitivity for bosons with masses $1\,{\rm eV} \lesssim m_X \lesssim 1\,\rm{MeV}$ that thermalize in the early Universe with neutrinos. Both of these experiments are fully funded and expected to probe these regions of parameter space within a decade. 

\section*{Acknowledgments}
SJW acknowledges support through the program Ram\'{o}n y Cajal (RYC2021-030893-I) of the Spanish Ministry of Science and Innovation, and through the European Research Council (ERC) under the European Union’s Horizon 2020 research and innovation programme (Grant agreement No. 864035 – Undark) and the Netherlands eScience Center, grant number ETEC.2019.018.
The work of SS received the support of a fellowship from “la Caixa” Foundation (ID 100010434) with fellowship code LCF/BQ/DI19/11730034.
SS also thanks the CERN theory group, the Lawrence Berkeley National Laboratory and the Berkeley Center for Theoretical Physics for hospitality.
We gratefully acknowledges the computer resources at Artemisa, funded by the European Union ERDF and Comunitat Valenciana as well as the technical support provided by the Instituto de Fisica Corpuscular, IFIC (CSIC-UV).

\newpage 

\bibliography{CMB_Constraints_on_eVBosons}

\maketitle
\onecolumngrid
\newpage
\begin{center}
\vspace{0.05in}
{ \it \large Supplementary Material for Precision CMB constraints on eV-scale bosons coupled to neutrinos
}\\ 
\vspace{0.05in}
{}
\end{center}
\onecolumngrid
\setcounter{equation}{0}
\setcounter{section}{0}
\setcounter{table}{0}
\makeatletter
\renewcommand{\theequation}{S\arabic{equation}}
\renewcommand{\thefigure}{S\arabic{figure}}
\renewcommand{\thetable}{S\arabic{table}}


In the supplementary material we provide additional information on the equations governing the evolution of the number density and energy density in the neutrino and bosonic fluids, and discuss the modifications made to CLASS. 
We also provide additional plots to illustrate the evolution of the background, the effect of the damping of neutrino free-streaming on the perturbations, and the impact of varying the mass and coupling on the temperature and polarization power spectra.

\section{Evolution of the background}
\label{SM:theory}
We use the formalism developed in~\cite{Escudero:2018mvt, EscuderoAbenza:2020cmq} to trace the evolution of the background, which assumes that the distribution functions for all relevant species can be characterized by their temperature $T_i$ and chemical potential $\mu_i$. The time evolution equations for these quantities reads
\begin{subequations}
\label{eq:full_system_maj}
\begin{align}
\frac{dT_\nu}{dt} &=\frac{1}{\frac{\partial n_\nu}{\partial \mu_\nu} \frac{\partial \rho_\nu}{\partial T_\nu}-\frac{\partial n_\nu}{\partial T_\nu} \frac{\partial \rho_\nu}{\partial \mu_\nu} }\left[ -3 H  \left((p_\nu+\rho_\nu)\frac{\partial n_\nu}{\partial \mu_\nu}-n_\nu \frac{\partial \rho_\nu}{\partial \mu_\nu} \right)+ \frac{\partial n_\nu}{\partial \mu_\nu}  \frac{\delta \rho_\nu}{\delta t} - \frac{\partial \rho_\nu}{\partial \mu_\nu}  \frac{\delta n_\nu}{\delta t} \right] , 
 \\
\frac{d\mu_\nu}{dt} &=\frac{-1}{\frac{\partial n_\nu}{\partial \mu_\nu} \frac{\partial \rho_\nu}{\partial T_\nu}-\frac{\partial n_\nu}{\partial T_\nu} \frac{\partial \rho_\nu}{\partial \mu_\nu} } \left[ -3 H \left((p_\nu+\rho_\nu)\frac{\partial n_\nu}{\partial T_\nu}-n_\nu \frac{\partial \rho_\nu}{\partial T_\nu} \right)+ \frac{\partial n_\nu}{\partial T_\nu}  \frac{\delta \rho_\nu}{\delta t} - \frac{\partial \rho_\nu}{\partial T_\nu} \frac{\delta n_\nu}{\delta t} \right]   , \\
\frac{dT_X}{dt} &=\frac{1}{\frac{\partial n_X}{\partial \mu_X} \frac{\partial \rho_X}{\partial T_X}-\frac{\partial n_X}{\partial T_X} \frac{\partial \rho_X}{\partial \mu_X} }\left[ -3 H  \left((p_X+\rho_X)\frac{\partial n_X}{\partial \mu_X}-n_X \frac{\partial \rho_X}{\partial \mu_X} \right)+ \frac{\partial n_X}{\partial \mu_X}  \frac{\delta \rho_X}{\delta t} -  \frac{\partial \rho_X}{\partial \mu_X}  \frac{\delta n_X}{\delta t} \right] , 
 \\
\frac{d\mu_X}{dt} &=\frac{-1}{\frac{\partial n_X}{\partial \mu_X} \frac{\partial \rho_X}{\partial T_X}-\frac{\partial n_X}{\partial T_X} \frac{\partial \rho_X}{\partial \mu_X} } \left[ -3 H \left((p_X+\rho_X)\frac{\partial n_X}{\partial T_X}-n_X \frac{\partial \rho_X}{\partial T_X} \right)+ \frac{\partial n_X}{\partial T_X}  \frac{\delta \rho_X}{\delta t} - \frac{\partial \rho_X}{\partial T_X} \frac{\delta n_X}{\delta t} \right]  ,  
\end{align}
\end{subequations}
where in these expressions $\rho_i$, $n_i$, and $p_i$ are the energy density, number density, and pressure of the given species $i$. 
The change in number and energy density of $X$ bosons per unit time,  $\delta n_X/\delta t $ and $\delta \rho_X/\delta t$, are given by
\begin{subequations}\label{eq:dndrho_dt_maj}
\begin{align}
\frac{\delta n_X}{\delta t} &= N_{\rm int}\, \Gamma(X \to \bar{\nu}\nu)\, \frac{m_X^2 }{2 \pi ^2} \times \left[T_\nu e^{\frac{2 \mu_\nu }{T_\nu}} K_1\left(\frac{m_X}{T_\nu}\right)-T_X e^{\frac{\mu_X}{T_X}} K_1\left(\frac{m_X}{T_X}\right)\right]\,, \\ 
\label{eq:deltarho_deltat}
\frac{\delta \rho_X}{\delta t} &= N_{\rm int}\, \Gamma(X \to \bar{\nu}\nu)\, \frac{m_X^3}{2 \pi ^2} \times\left[T_\nu e^{\frac{2 \mu_\nu }{T_\nu}} K_2\left(\frac{m_X}{T_\nu}\right)-T_X e^{\frac{\mu_X}{T_X}} K_2\left(\frac{m_X}{T_X}\right)\right]\,.
\end{align}
\end{subequations}
Here we have used the Maxwell-Boltzmann approximation, and introduced the Bessel functions $K_i$. 
Since the process we are considering is $1\to 2$, the transfer rate for neutrinos are related to these shown here via: $\delta\rho_\nu/\delta t = -\delta\rho_X/\delta t$ and $\delta n_\nu/\delta t = -2\delta n_X/\delta t$.
For photons and neutrinos which are not coupled to the light neutrino-philic boson, one simply has $\frac{dT}{dt} = -H \, T$.

The evolution equations shown above, along with the modified perturbation equations outlined in Eqns.~\eqref{eq:pert_delta}-\eqref{eq:Fl}, are implemented into the publicly available cosmological Boltzmann solver CLASS~\cite{Blas:2011rf}. The incorporation of the new interactions into \texttt{CLASS} follows the standard methodology and is done in three steps.
First, the \texttt{input} module is modified to read the new model parameters, including the coupling constant $\lambda_\nu$, the mass of the new boson $m_X$, and its primordial abundance parametrized via $\Delta N_{\rm{eff}}^{\rm{BBN}}$. 
A list with a detailed description of all readable parameters can be found in the file \textit{majoron.ini} in the \gitlink\, repository of our code.
The background evolution of the system as governed by Eq.~\eqref{eq:full_system_maj} is implemented into the function \texttt{int background\_derivs} within the \texttt{background} module.
Its initial conditions are defined in the function \texttt{int background\_initial\_conditions}.
Because the differential equations governing the background evolution are generally stiff, we explicitly implement a reduction of the step size when the energy density in the new species is non-negligble -- specifically, we take this range to be $0.65 < T_\nu/m_X < 10$.
The step size can be controlled in the \textit{.ini} file via the parameter \texttt{fine\_steps\_maj} and is typically $\mathcal{O}(10^{-5})$.
The evolution of all relevant background quantities can be accessed by different modules through the pointer \texttt{pba$\to$}.
This is of particular importance to calculate the perturbations in the \texttt{perturbations} module.
By accessing the background evolution, it is then straightforward to implement Eqns.~\eqref{eq:pert_delta} - \eqref{eq:Fl} into the function \texttt{int perturb\_derivs}.


\section{Impact on Cosmology}
\label{SM:cosmo}
Here, we take the opportunity to provide a more detailed picture of how neutrino-philic bosons alter the temperature and polarization power spectrum.
We begin by showing the evolution of the energy density stored in both neutrinos and the $X$ boson for scenarios with different number of interacting neutrino species, $N_{\rm{int}}$, scalar or vector $X$-bosons, and with and without primordial $X$ abundance.
We then discuss the  impact on the cosmological perturbations.
In particular, we show the evolution of the speed of sound $c_s^2$, equation of state $\omega$, the neutrino density contrast in the synchronous gauge $\delta_\nu$, and the neutrino anisotropic stress $\sigma_\nu$, and finally the impact on the temperature and polarization power spectra $C_\ell^{TT}$ and $C_\ell^{EE}$.

\begin{figure}[!ht]
\centering
\begin{tabular}{cc}
\hspace{-0.2cm} \includegraphics[width=0.47\textwidth]{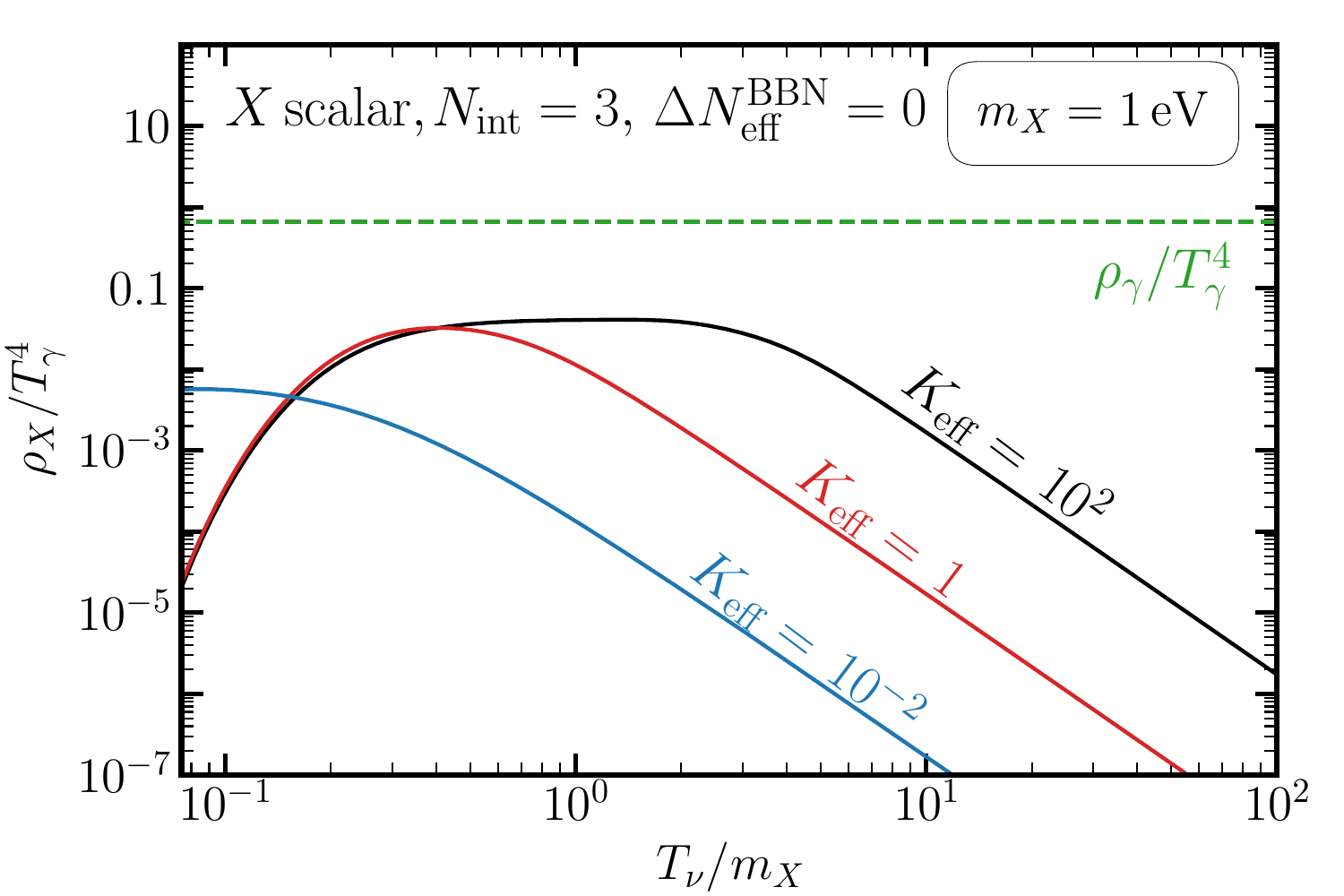} &
\hspace{-0.2cm} \includegraphics[width=0.47\textwidth]{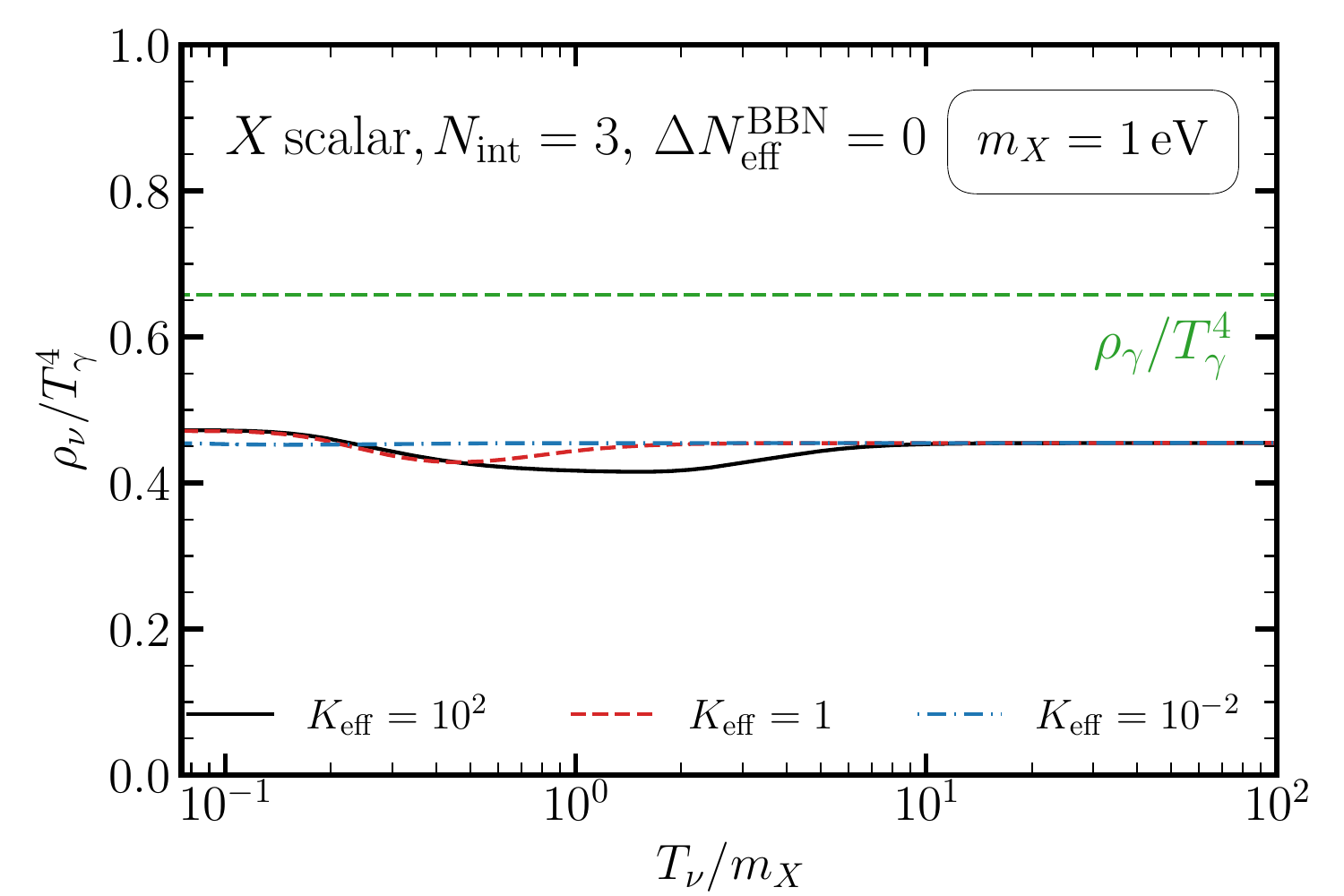}  \\
\hspace{-0.2cm} \includegraphics[width=0.47\textwidth]{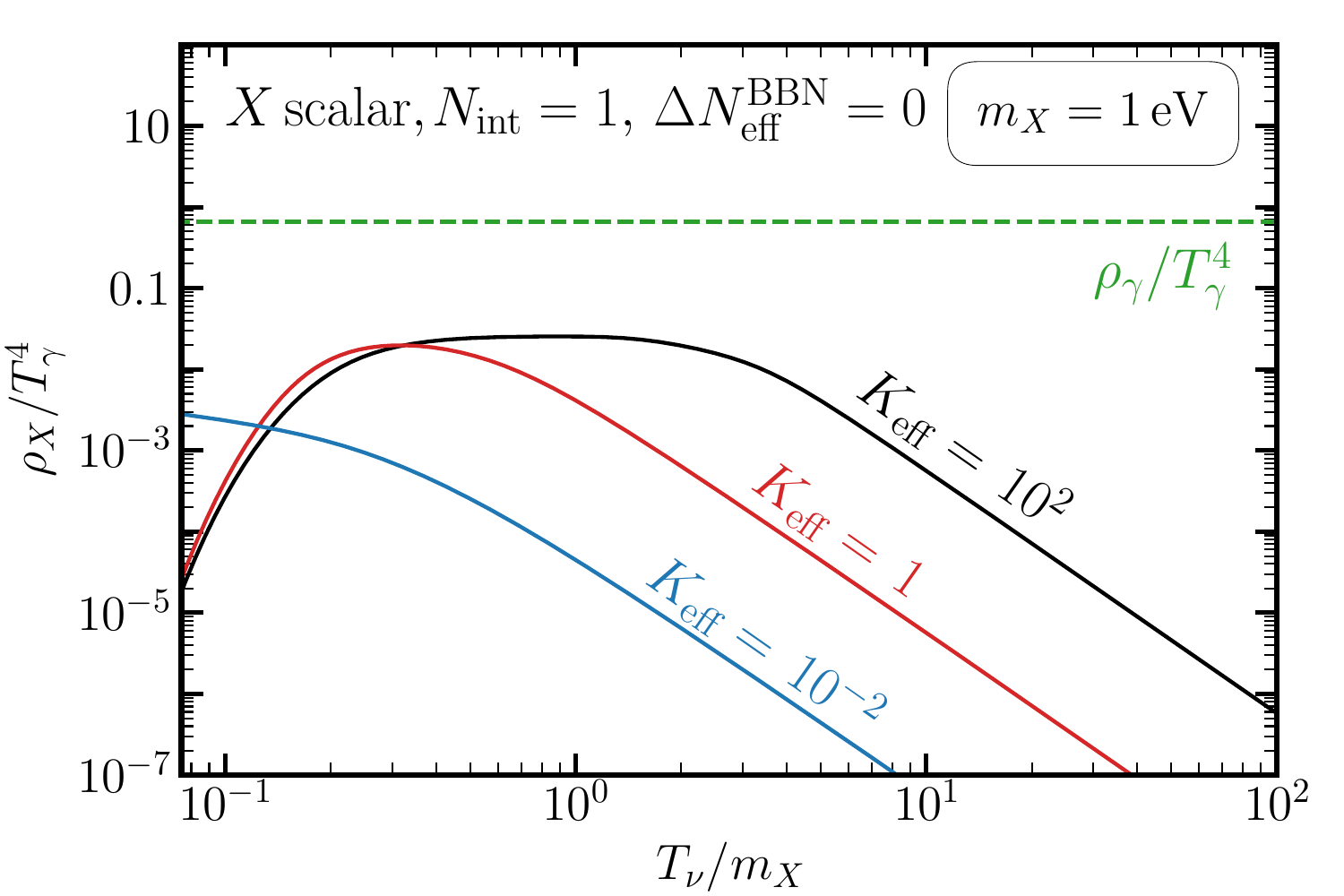} &
\hspace{-0.2cm} \includegraphics[width=0.47\textwidth]{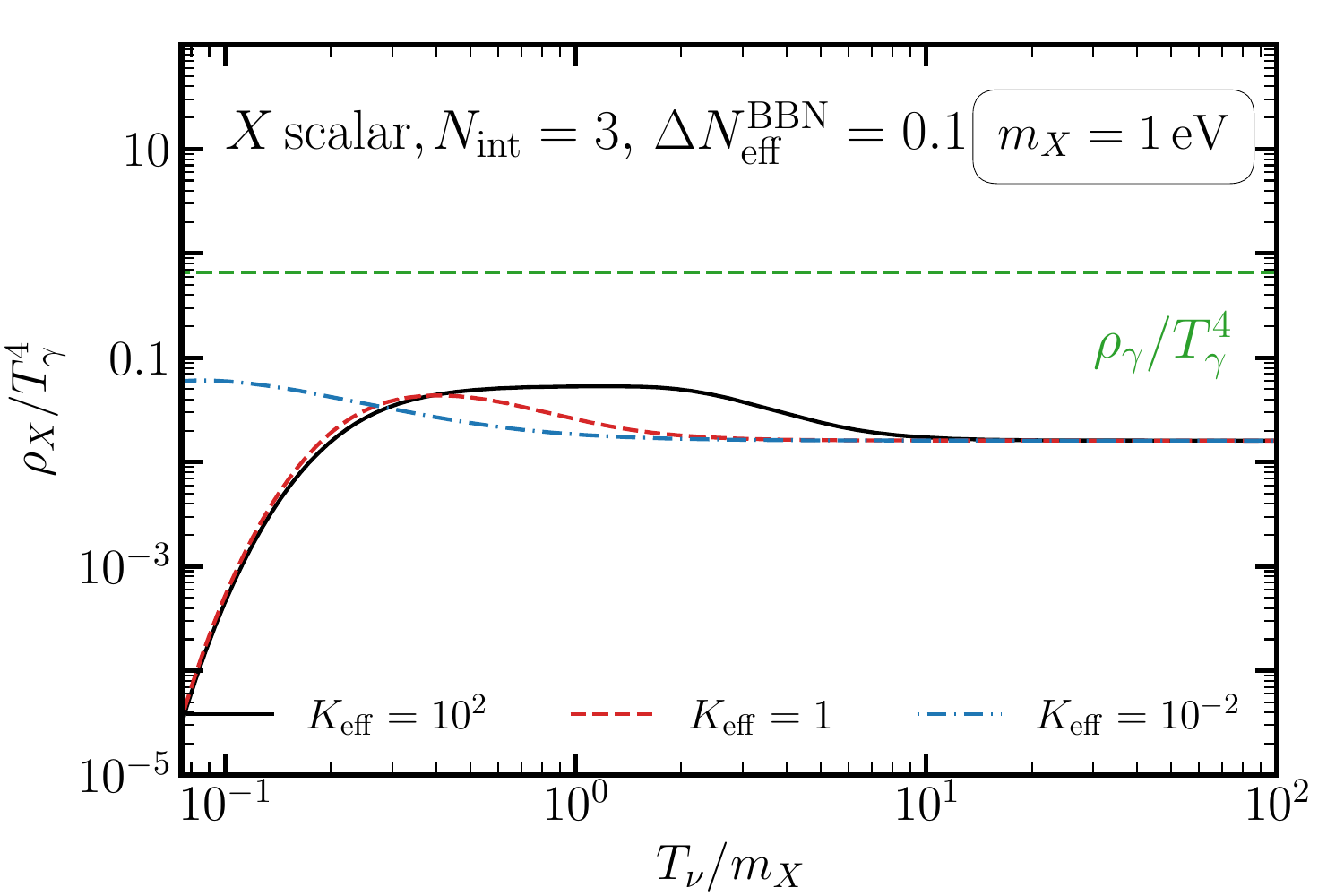}  \\
\hspace{-0.2cm} \includegraphics[width=0.47\textwidth]{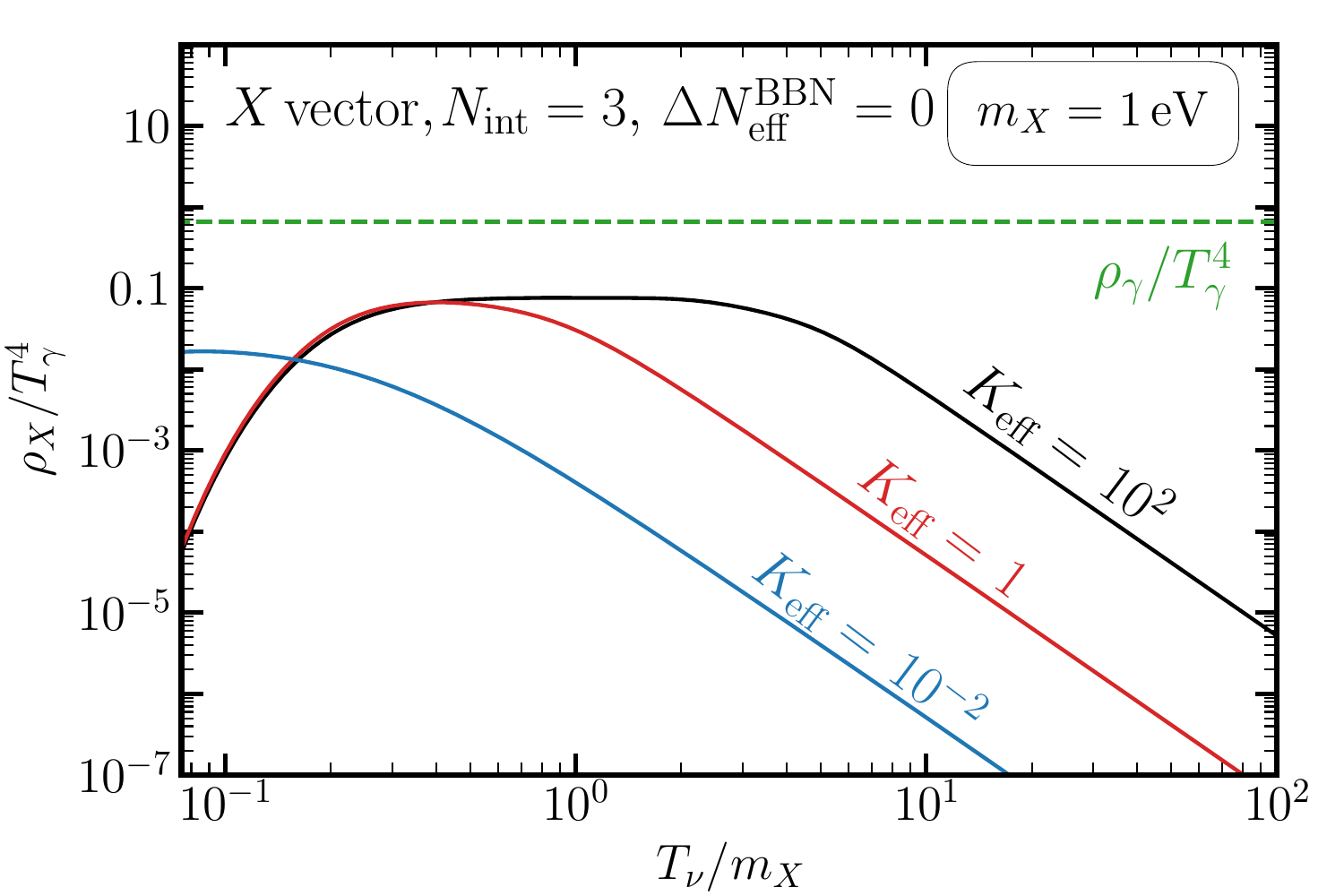} &
\hspace{-0.2cm} \includegraphics[width=0.47\textwidth]{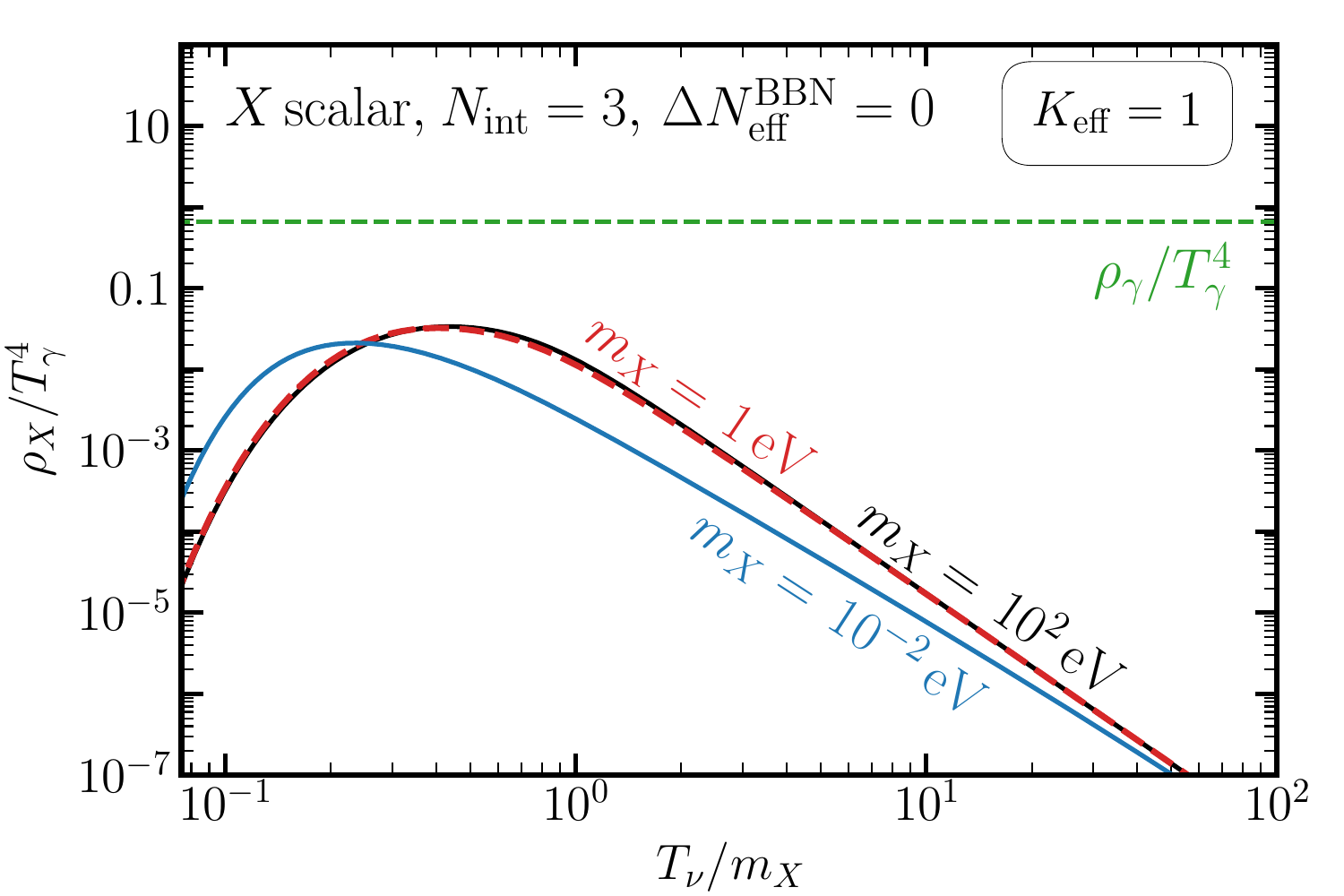}  
\end{tabular}
\vspace{-0.4cm}
\caption{
Evolution of the energy density of the neutrino-philic boson $X$ as well as the neutrinos for a variation of different cases as specified in the plot labels. 
}
\label{fig:background_detail}
\end{figure}

In figure~\ref{fig:background_detail} we show the evolution of the energy density of the neutrino and $X$ boson for various boson masses and interaction strengths, varying the primordial abundances, the number of interacting neutrinos, and the spin of the neutrino-philic boson. 
The general shape of the energy density evolution of the $X$ boson is the same as long as $\Delta N_{\rm{eff}}^{\rm{BBN}} = 0$.
This can be seen from the left column of Figure~\ref{fig:background_detail}.
The only difference between the cases with different number of interacting neutrino families and $X$ being a scalar or vector boson is the peak density the $X$ boson reaches (while keeping the other model parameters fixed).
In general, the more neutrino families interact with the $X$ particles, the more the $X$ particle will be populated in the thermal plasma.
The same holds for the vector versus scalar case, i.e. $\rho_X^{\rm{vector}} > \rho_X^{\rm{scalar}}$ at the peak of its thermalization history. 
Additionally, in the bottom right panel of Figure~\ref{fig:background_detail} we show the evolution of the $X$ particle density for fixed $K_{\rm{eff}} = 1$ with varying mass $m_X$.
We see that $\rm{max}(\rho_X)$ is approximately the same for all depicted masses.
On the other hand, if e.g. $\Delta N_{\rm{eff}}^{\rm{BBN}} = 0.1$, the relative change of the energy density between its initial value and its maximal value is relatively small.
However, the striking feature is that for $K_{\rm{eff}} \ll 1$ the boson energy density, $\rho_X$, reaches larger values at lower temperature compared to the $K_{\rm{eff}} \gg 1$ scenario.
This is contrary to the case of vanishing primordial $X$ particle abundance.
The reason is, as outlined in Section~\ref{sec:formalism}, that the X boson becomes non-relativistic and its small coupling leads to a delayed out-of-equilibrium decay.
Lastly, in the top right panel of Figure~\ref{fig:background_detail} we also show a exemplary evolution of the neutrino energy density $\rho_\nu$ for the case of $X$ being of scalar type with vanishing primordial abundance and interacting with all three neutrino families.
We choose to fix $m_X = 1\,\rm{eV}$ and vary $K_{\rm{eff}}$.
This makes it evident that the energy density stored in the neutrinos is enhanced for $K_{\rm{eff}} \gtrsim 1$ compared to weakly and non-interacting neutrino scenarios.
On the other hand, $\rho_\nu(T_\nu \to 0)$ also saturates to a maximal value and becomes independent of $K_{\rm{eff}}$ as long as $K_{\rm{eff}} \gtrsim 1$.
We can also translate the evolution of these energy densities into $N_{\rm{eff}}$, see Figure~\ref{fig:Neff_example}, via
\begin{align}
    N_{\rm{eff}} \equiv \frac{8}{7} \left( \frac{11}{3} \right)^{4/3} \left( \frac{\rho_{\rm{rad}} - \rho_\gamma}{\rho_\gamma} \right)\,,
\end{align}
where $\rho_{\rm{rad}}$ is the total energy density stored in radiation.
This directly relates the late time enhancement of the energy density in the neutrinos and $X$ boson to the observable measured by Planck.

\begin{figure}[!ht]
\centering
\begin{tabular}{cc}
\hspace{-0.2cm} \includegraphics[width=0.47\textwidth]{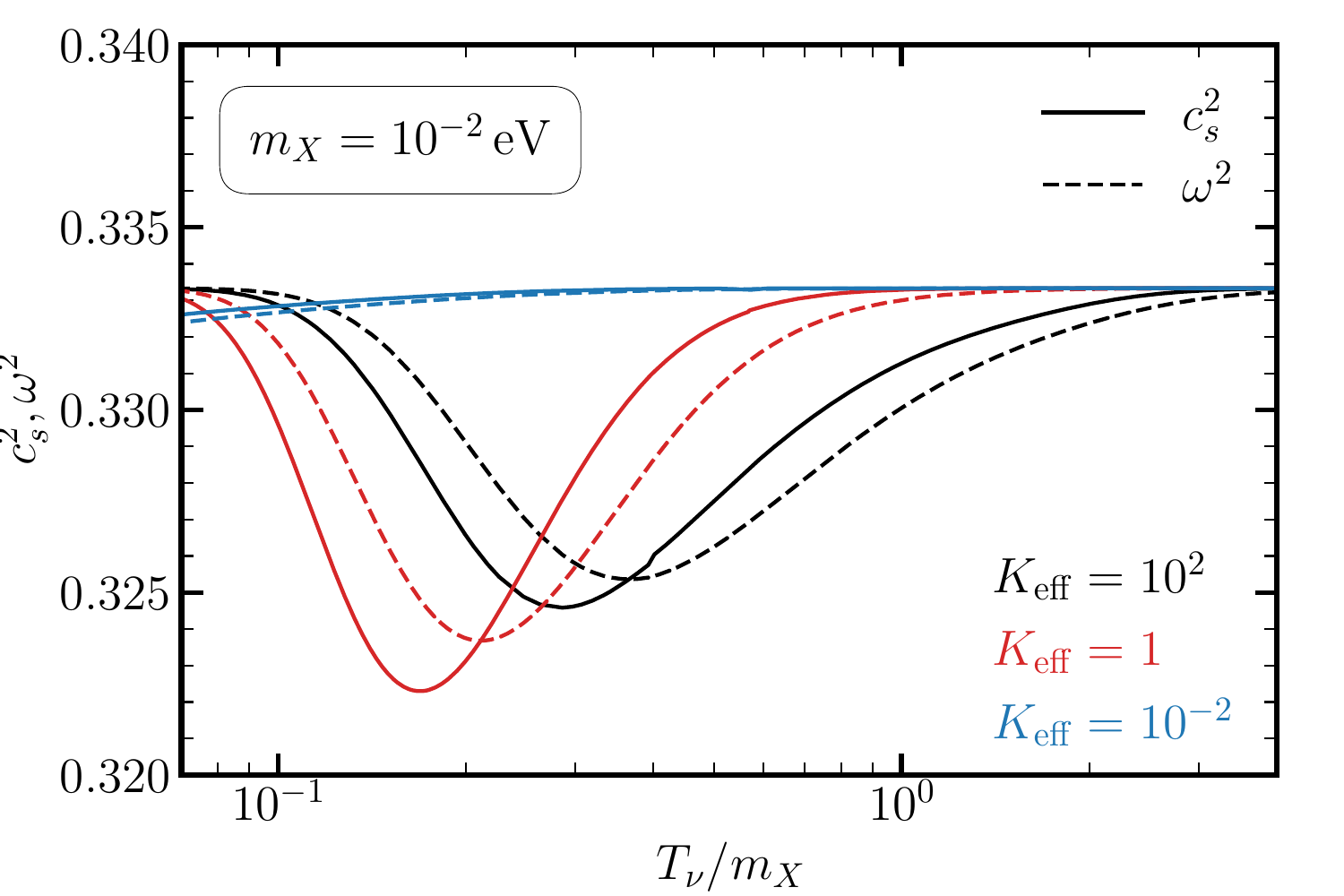} &
\hspace{-0.2cm} \includegraphics[width=0.47\textwidth]{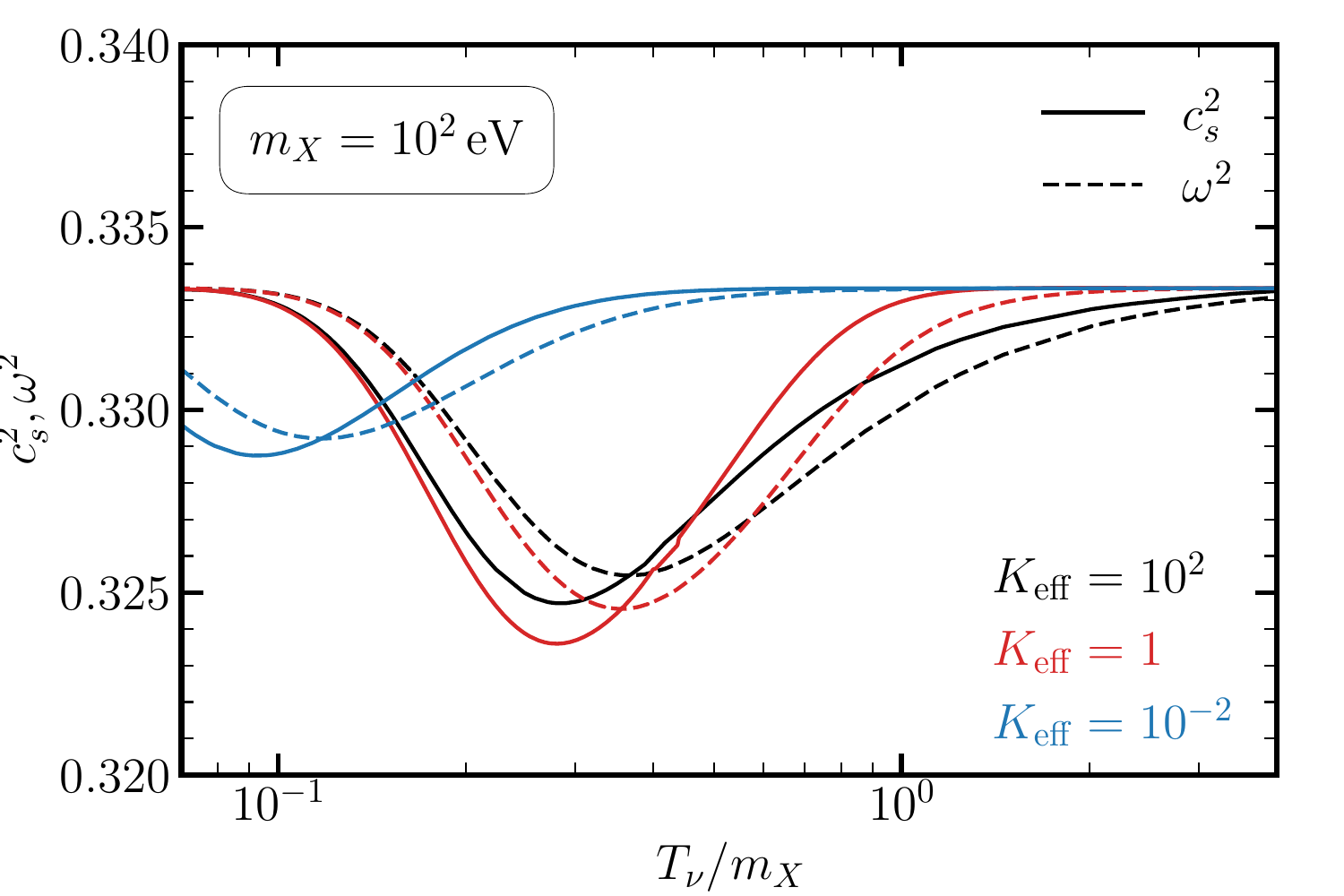}  \\
\end{tabular}
\vspace{-0.4cm}
\caption{Speed of sound and equation of state of the joint neutrino+$X$ boson system.
A fully relativistic fluid will have $c_s^2 = \omega = 1/3$. 
}
\label{fig:SM_cs_w}
\end{figure}

The solutions derived from solving Eq.~\ref{eq:full_system_maj} also allow to compute the sound speed, $c_s^2$, and the equation of state, $\omega$, of the joint fluid.
The solution for two different masses of the neutrino-philic boson and different interactions strengths are shown in figure~\ref{fig:SM_cs_w}.
We see that the maximal deviation from the relativistic approximation $c_s^2 = \omega = 1/3$ is reached for $K_{\rm{eff}} = 1$, independent of the mass of the $X$ particle $m_X$.
In particular, we can appreciate that also for $m_X \ll 1\,\rm{eV}$ the evolution significantly deviates from the relativistic approximation.
Although differences of $\mathcal{O}(4 \%)$ can arise with respect to the relativistic approximation, we explicitly checked that this leads to a negligible effect on all observables.
This approximation typically induces an error in the TT power spectrum at the level of $\mathcal{O}(0.01\%)$ in all relevant regions of parameter space and is therefore well below Planck sensitivity.

\begin{figure}[!ht]
\centering
\begin{tabular}{cc}
\hspace{-0.2cm} \includegraphics[width=0.47\textwidth]{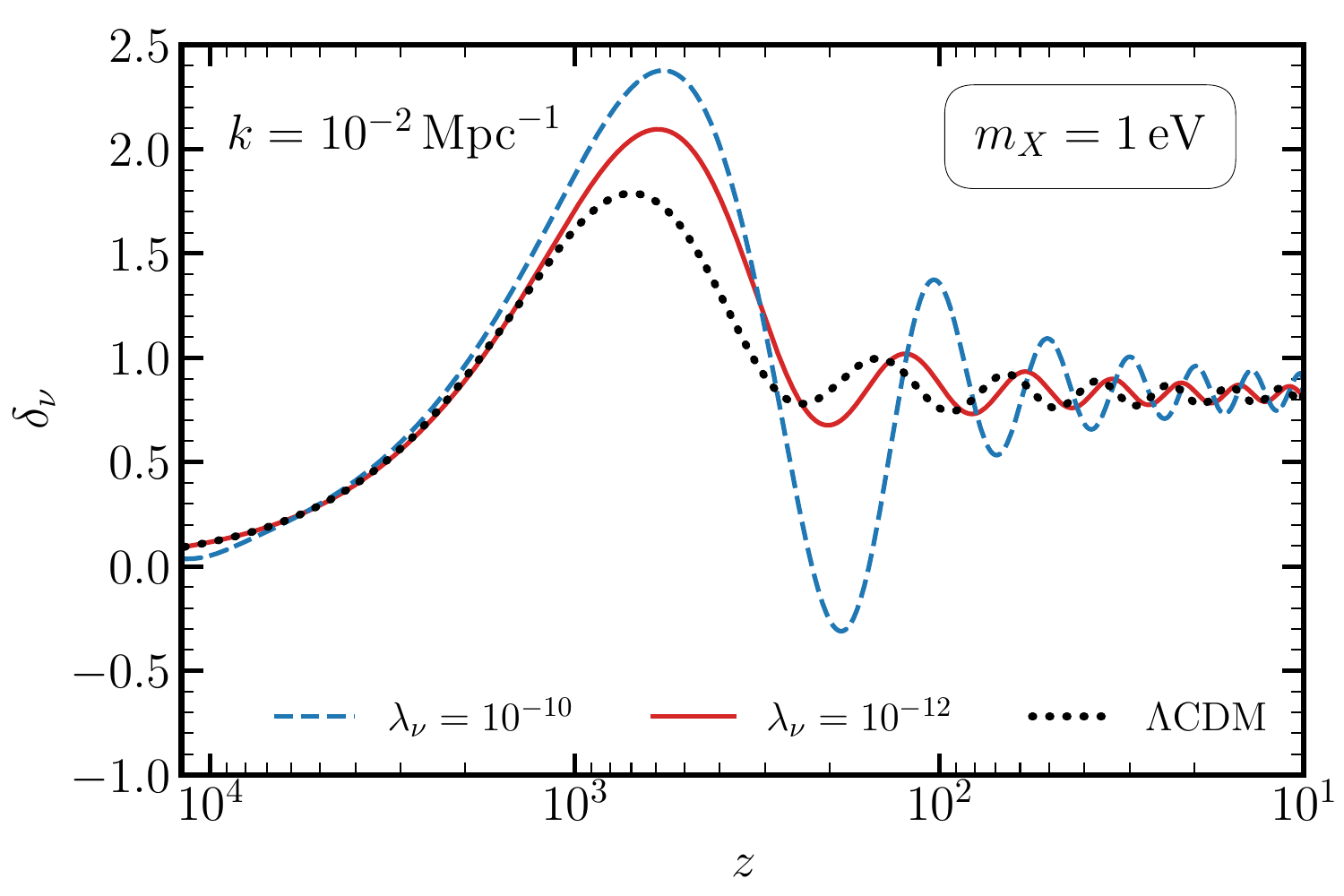} &
\hspace{-0.2cm} \includegraphics[width=0.47\textwidth]{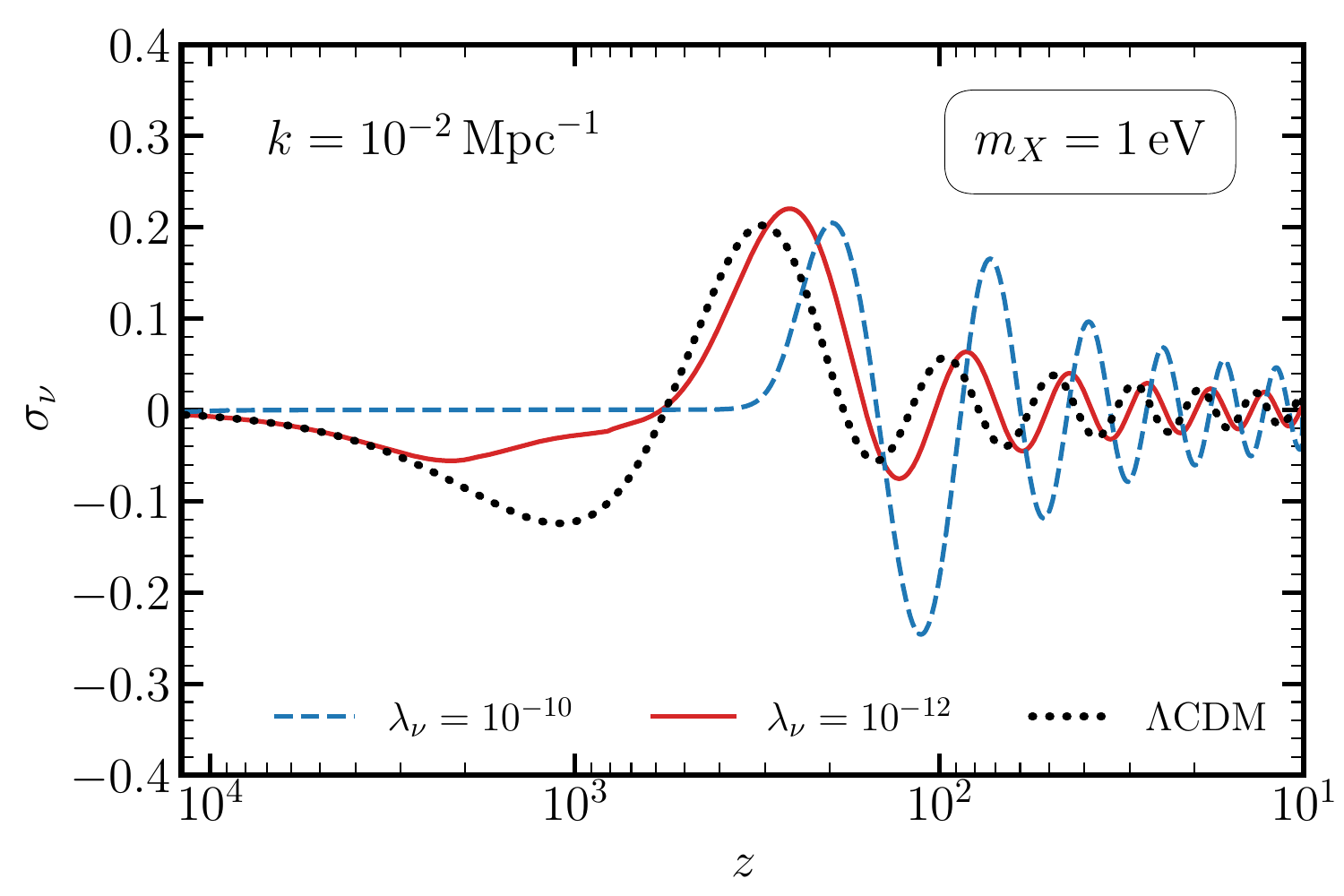}  \\
\hspace{-0.2cm} \includegraphics[width=0.47\textwidth]{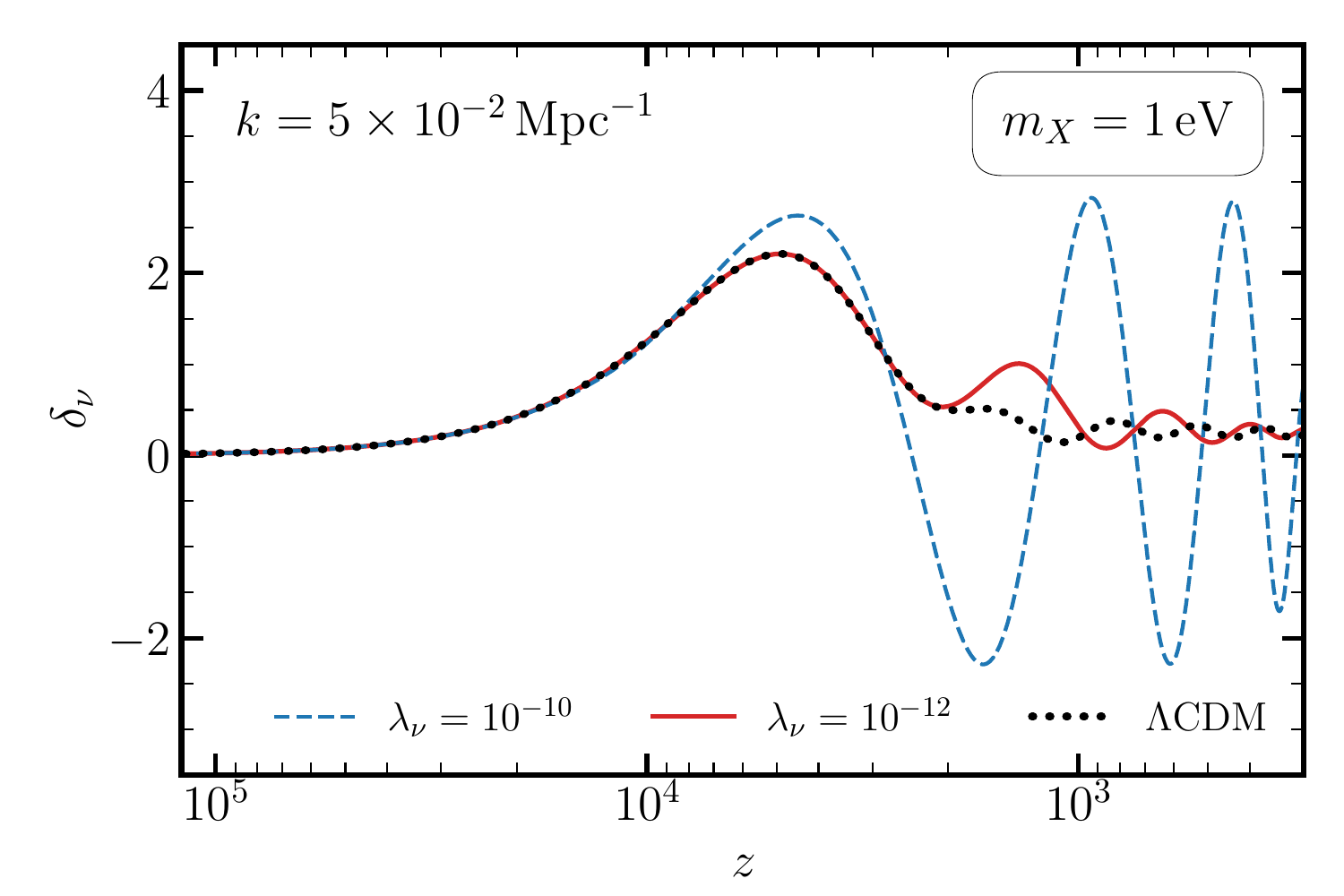} &
\hspace{-0.2cm} \includegraphics[width=0.47\textwidth]{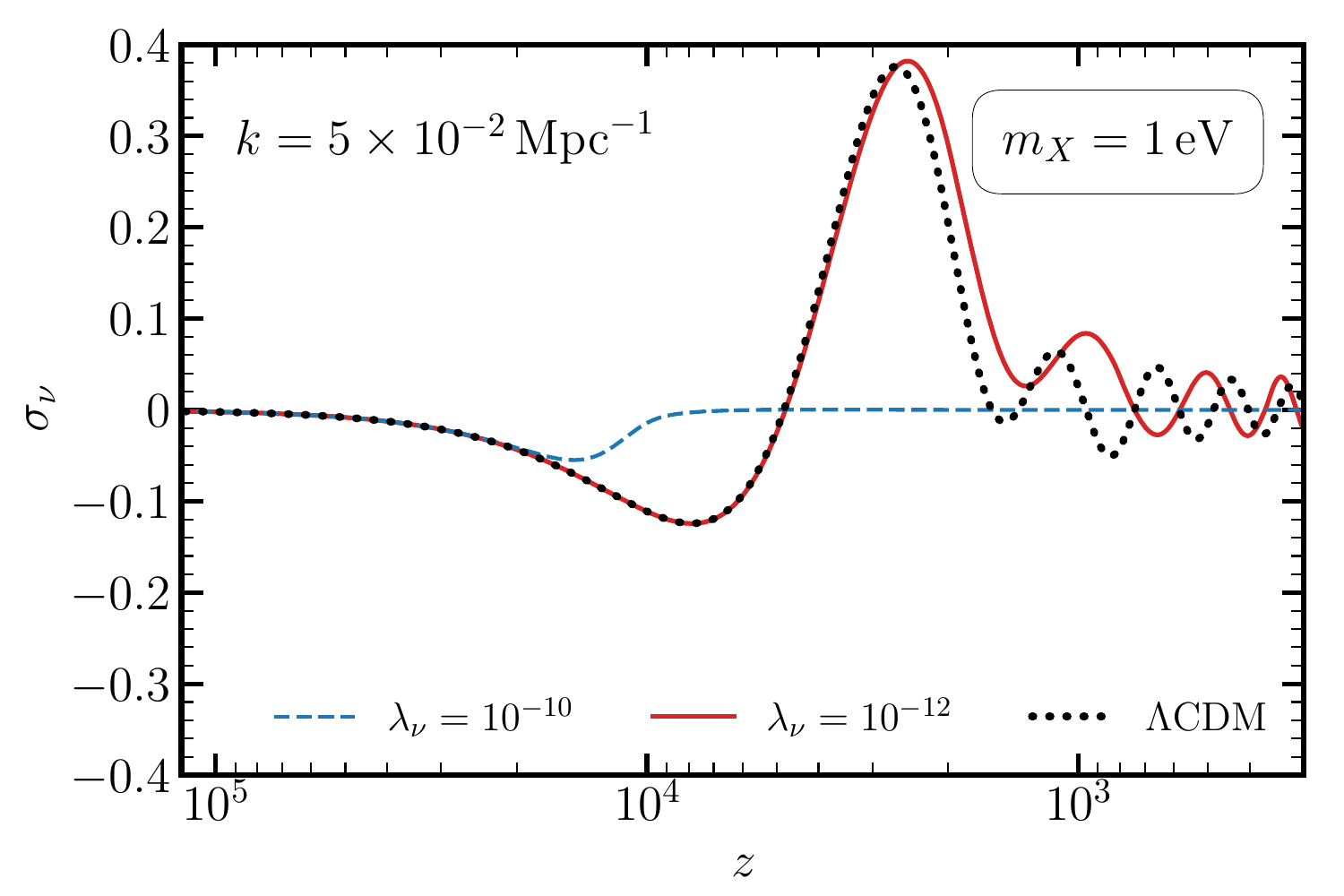}  
\end{tabular}
\vspace{-0.4cm}
\caption{
Density contrast $\delta_\nu$ and the neutrino anisotropic stress (shear) $\sigma_\nu$ in synchronous gauge for fixed mass $m_X = 1\,\rm{eV}$, but different wavelengths $k$ and interaction strengths $\lambda_\nu$.
}
\label{fig:SM_delta_shear}
\end{figure}

We also show the evolution of the density contrast $\delta_\nu$ and the neutrino anisotropic stress (shear) $\sigma_\nu$ for two different wavelengths in Figure~\ref{fig:SM_delta_shear}.
For reference, the black dotted lines indicate the $\Lambda$CDM expectation.
The neutrino interactions lead to a significant reduction of the neutrino anisotropic stress at the time of recombination $z \sim 10^{3}$, see also Figure~\ref{fig:CLs_shear}.
On the other hand, the very same interaction leads to a significant enhancement of the density contrast at the same time.

\begin{figure}[!ht]
\centering
\begin{tabular}{cc}
\hspace{-0.2cm} \includegraphics[width=0.47\textwidth]{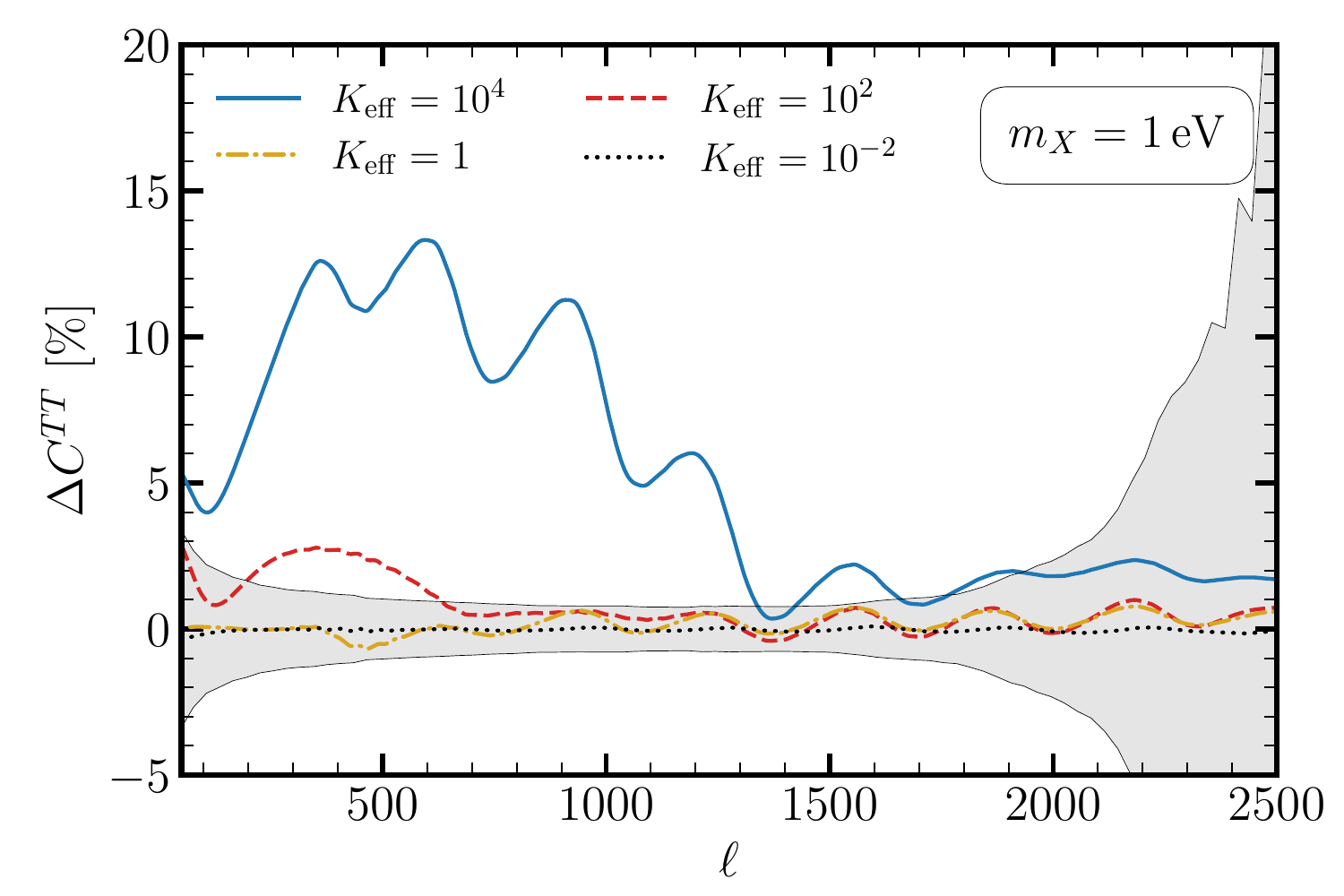} &
\hspace{-0.2cm} \includegraphics[width=0.47\textwidth]{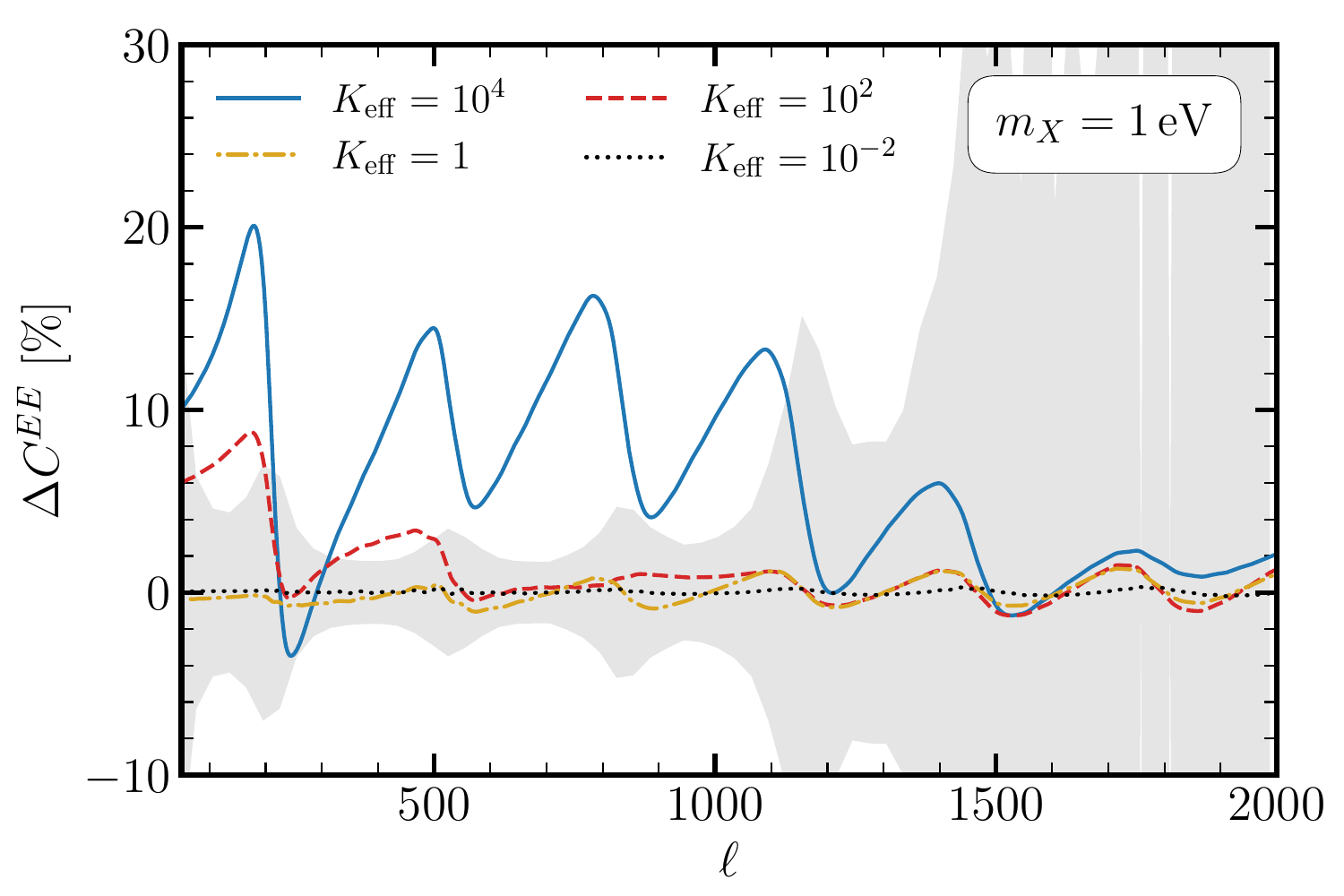} 
\end{tabular}
\vspace{-0.4cm}
\caption{
Fractional difference on the TT (EE) power spectrum with respect to $\Lambda$CDM for the case of a scalar particle interacting with neutrinos in the left (right) panel. 
We show the results for fixed mass $m_X = 1\,\rm{eV}$ and different values of $K_{\rm{eff}}$, see Eq.~\eqref{eq:Keff_param}.
}
\label{fig:SM_Cl}
\end{figure}

Finally, we show the variation of the temperature fluctuation $C_\ell^{TT}$ and polarization $C_\ell^{EE}$ spectrum in the left and right panel of Figure~\ref{fig:SM_Cl}.
The temperature polarization spectrum shown here is complementary to the one in Figure~\ref{fig:CLs_masschange} of the main text, in which we show the variation for fixed interactions strength and different masses $m_X$.
Here, we fix the mass to be $m_X = 1\,\rm{eV}$ and vary the interactions strength $K_{\rm{eff}}$.
As expected, interactions with strength $K_{\rm{eff}} \gg 1$ lead to a significant perturbation of both spectra, $C_\ell^{TT}$ and $C_\ell^{EE}$, which can exceed the $1\sigma$ error bars of the Planck mission shown in grey.
In particular, the interactions induce a periodic perturbation spectrum with strong damping for the high-$\ell$ multipole moments as dictated by Eq.~\eqref{eq:GNF}.

\begin{figure*}[t]
\centering
\begin{tabular}{cc}
\hspace{-0.5cm} \includegraphics[width=0.47\textwidth]{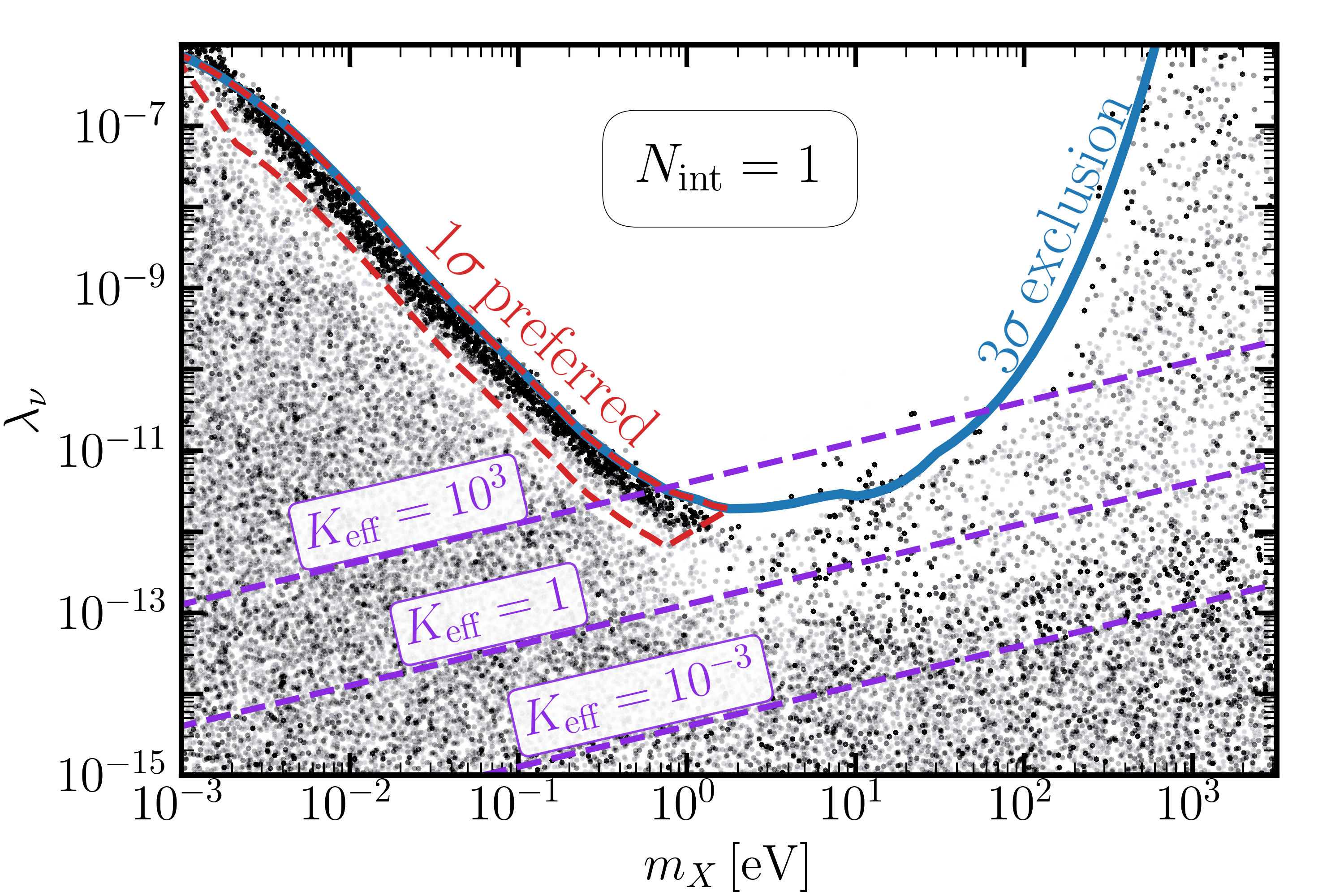} &
\hspace{-0.1cm} \includegraphics[width=0.47\textwidth]{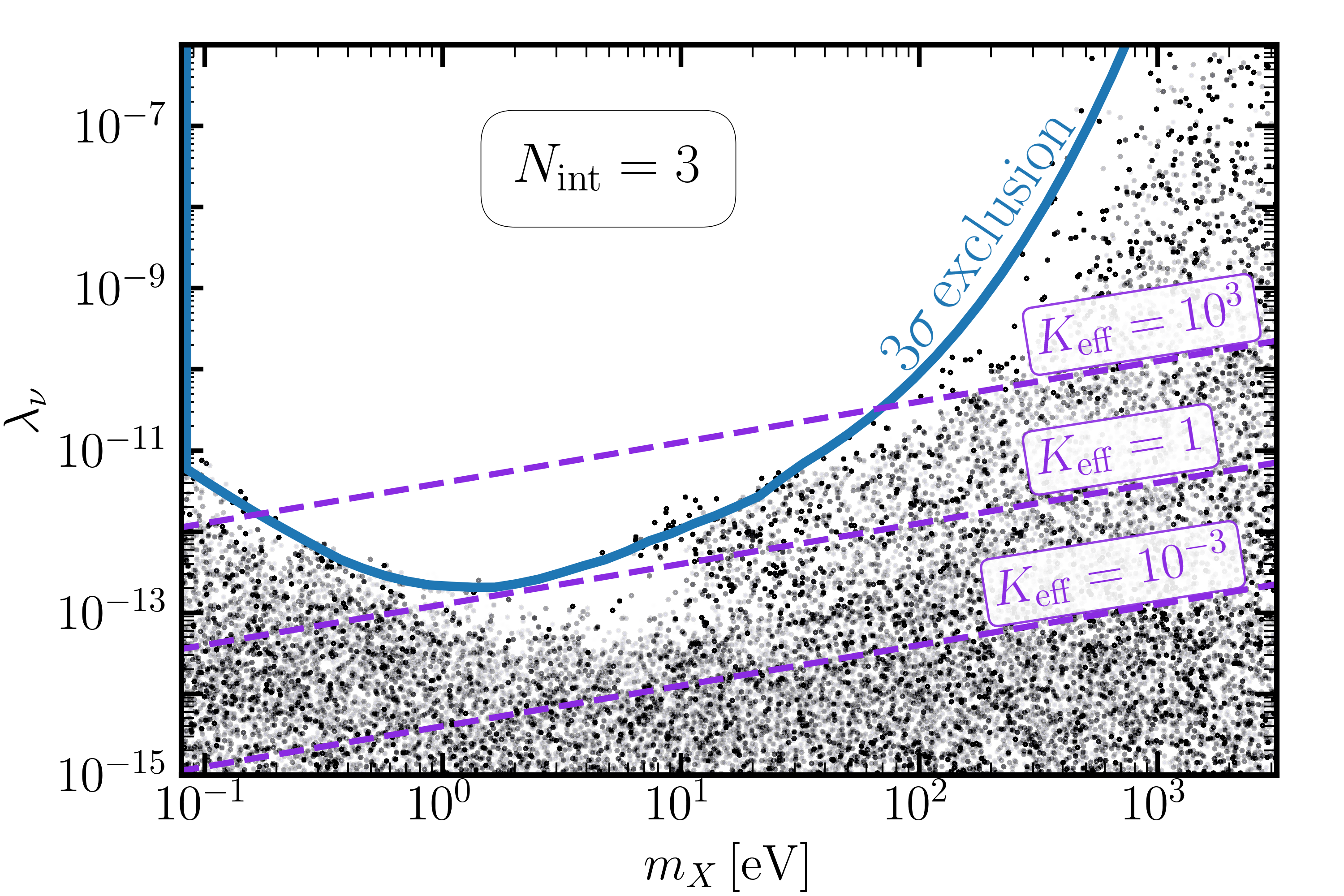} 
\end{tabular}
\vspace{-0.4cm}
\caption{
Results of a MCMC analysis against the Planck legacy data. 
Black scattered points correspond $N\sim 2\times 10^6$ Monte Carlo samples in the analysis. 
We find that $99.7\%$ of all points are below the blue line, labeled as $3\sigma$ exclusion.
In the left (right) panel we show the results for a scalar particle interacting with $N_{\rm int} = 1$ ($N_{\rm int} = 3$) neutrino families.
Interestingly, for the $N_{\rm int} = 1$ case the MCMC analysis identifies a $1\sigma$ preferred region, as can be seen by the red region.
This corresponds to a non-trivial clustering of points which furthermore exhibit significant neutrino-philic interactions.
In addition, for reference, we show in purple dashed the isocontours of fixed $K_{\rm eff}$. 
}
\label{fig:SM_PointsParameterspace}
\end{figure*}

\begin{figure}
    \centering
    \includegraphics[width=0.55\textwidth]{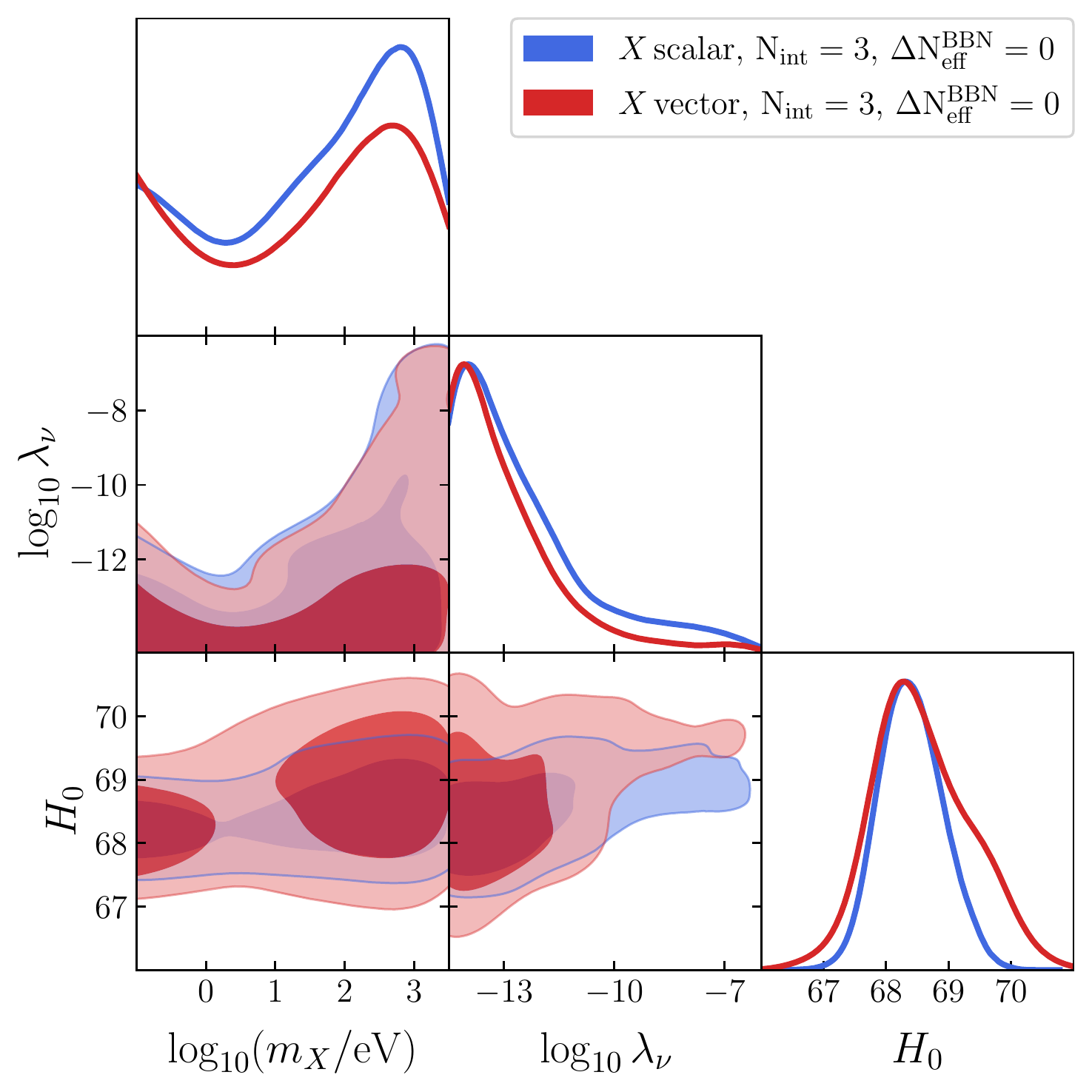}
    \caption{
    $1\sigma$ and $2\sigma$ posterior probabilities for the scenario of $N_{\rm{int}} = 3$, $\Delta N_{\rm{eff}}^{\rm{BBN}} = 0$.
    The case of $X$ being a scalar boson is shown in blue and the vector case is shown in red.
    The model is tested against the likelihood of the full Planck$18+$BAO data set.
    }
    \label{fig:SM_getdist_1}
\end{figure}

\begin{figure}
    \centering
    \includegraphics[width=0.55\textwidth]{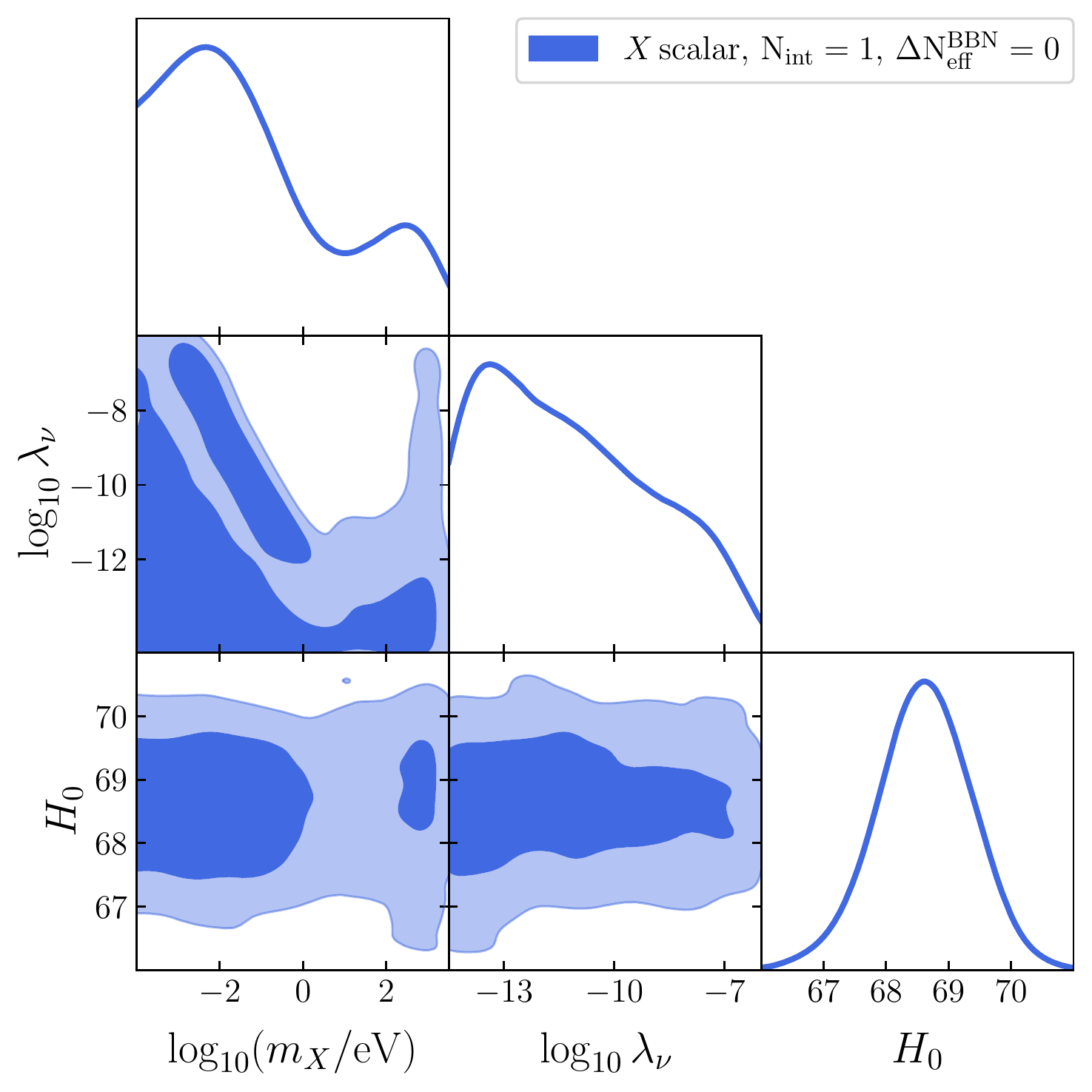}
    \caption{
    $1\sigma$ and $2\sigma$ posterior probabilities for the scenario of $N_{\rm{int}} = 1$, $\Delta N_{\rm{eff}}^{\rm{BBN}} = 0$ and $X$ being of scalar type.
    The model is tested against the likelihood of the full Planck$18+$BAO data set.
    }
    \label{fig:SM_getdist_2}
\end{figure}

\begin{figure}
    \centering
    \includegraphics[width=0.75\textwidth]{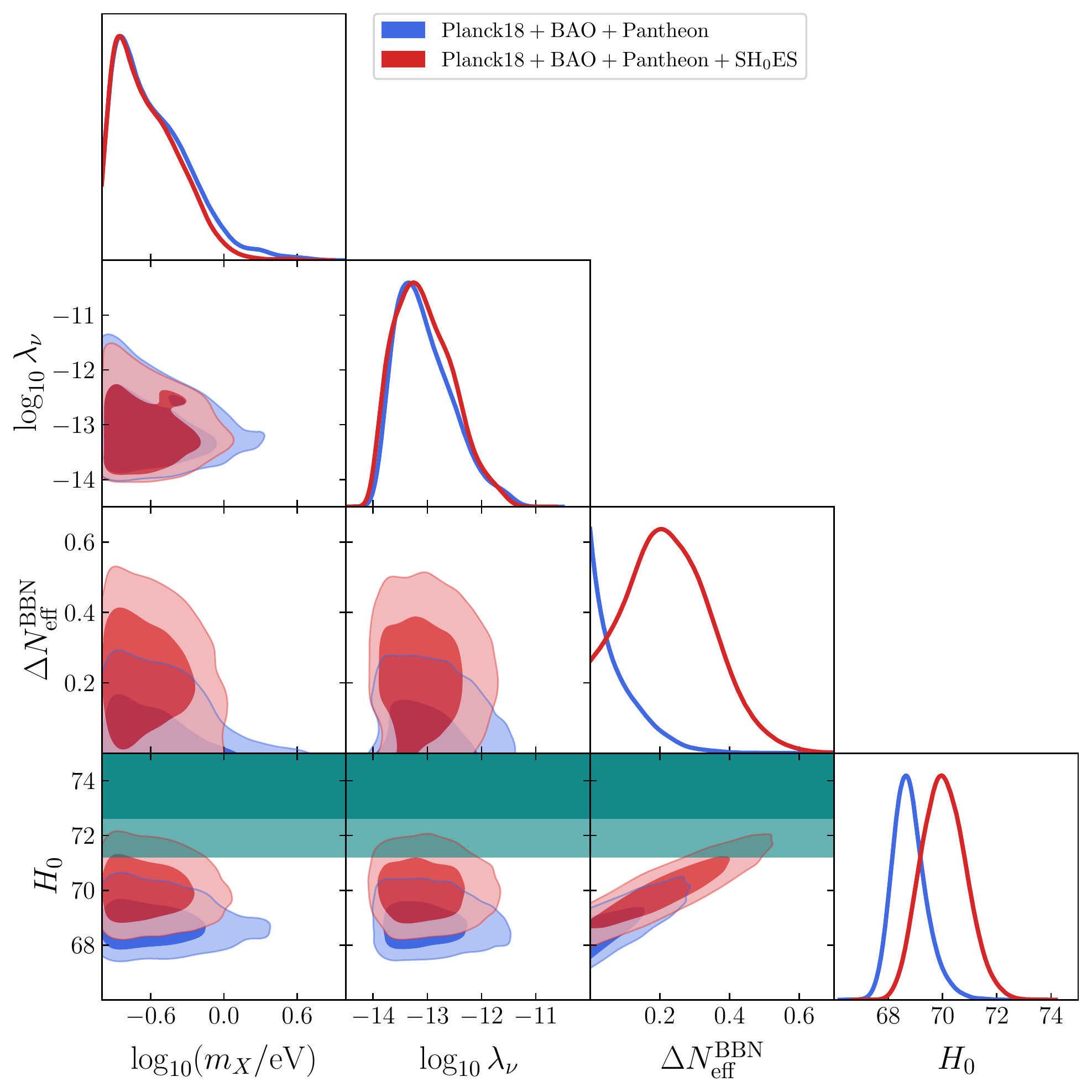}
    \caption{
    $1\sigma$ and $2\sigma$ posterior probabilities for the scenario of $N_{\rm{int}} = 1$ and $X$ being of scalar type.
    We explicitly allow for a non-vanishing primordial abundance of the $X$ particle, parametrized by $\Delta N_{\rm{eff}}^{\rm{BBN}}$.
    The model is tested against the likelihood of the full Planck$18+$BAO data set and adds i) the Pantheon data set (blue) and ii) the Pantheon$+\rm{SH_0ES}$ data sets (red).
    For reference we show in green the $1\sigma$ and $2\sigma$ $\rm{SH_0ES}$ posterior values.
    }
    \label{fig:SM_getdist_3}
\end{figure}

\begin{figure}
    \centering
    \includegraphics[width=0.99\textwidth]{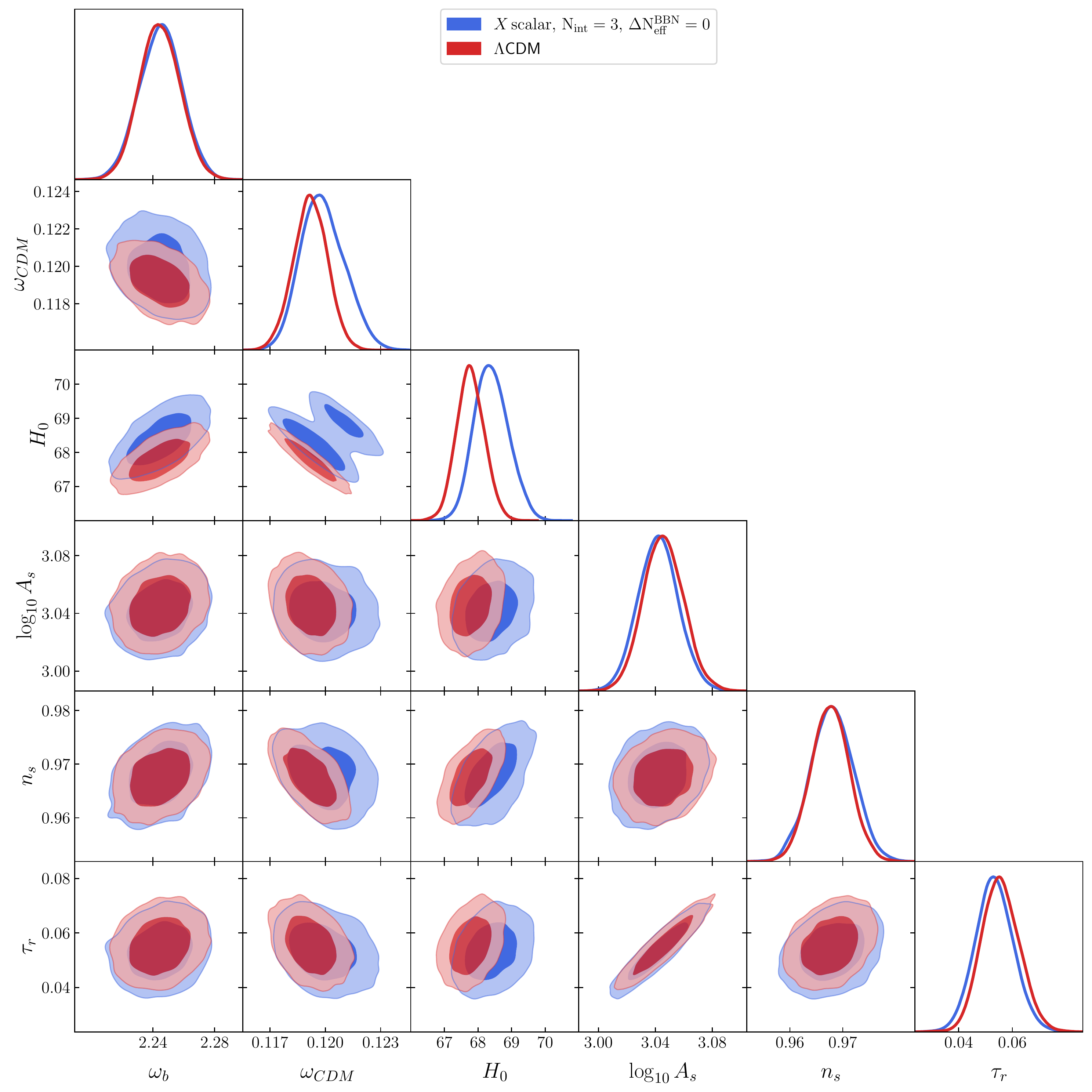}
    \caption{
    $1\sigma$ and $2\sigma$ posterior probabilities for the scenario of $N_{\rm{int}} = 3$, $\Delta N_{\rm{eff}}^{\rm{BBN}} = 0$ and $X$ being a scalar boson.
    We show the correlation of the standard cosmological parameters and compare to $\Lambda$CDM.
    }
    \label{fig:SM_getdist_4}
\end{figure}

Having implemented the background and perturbation differential equations for the joined neutrino-$X$ fluid, see Section~\ref{SM:theory} for details, an accurate parameter space scan can be done via a MCMC analysis.
We chose to implement our code into the publicly available software \texttt{MontePython}~\cite{Brinckmann:2018cvx, Audren:2012wb}.
The full analysis then leads to the main result as shown in Figure~\ref{fig:bounds_majoron}.
Here, we would like to give more specific details on two important aspects -- i) how we have identified the exclusion region and the best-fit region, and ii) the correlation of neutrino interactions with different cosmological parameters.
Let us start with point i).
After removing the non-markovian points as well as the burn-in points of each chain, the raw points projected onto the parameter space of $(m_X,\lambda_\nu)$ can be visualized as in Figure~\ref{fig:SM_PointsParameterspace}.
For clarity we choose to only depict the case of $X$ being a scalar mediator, but the same arguments also hold for the vector scenario.
The exclusion region is simply found by demanding that within a given mass bin $99.7\%$ of all points are below a given value of $\lambda_{\nu,i}$.
We chose a bin size of $\Delta \log_{10}(m_X/{\rm eV}) = 0.2$.
On the other hand, the best fit
region can be obtained by evaluating the cluster density of sampled points.
The same preferred region can also be found by running MCMC analysis softwares as e.g. GetDist~\cite{Lewis:2019xzd}.
This brings us to point ii) -- the triangle plots of different MCMC runs.
We analyze the chains with the publicly available GetDist software.
Of all the MCMC analysis done, we select a representative set of results.
\begin{itemize}
    \item Comparison between the case of $X$ being of scalar or vector type with fixed $\Delta N_{\rm{eff}}^{\rm{BBN}} = 0$ and $N_{\rm{int}} = 3$. The likelihood to be tested against is the full Planck $+$ BAO set.
    The result is shown in Figure~\ref{fig:SM_getdist_1}.
    The lower bound on the mass of $m_X$ is set by the requirement of $m_X > 2\times m_{\nu,i}$.
    We can see that for the $X$ vector case globally slightly lower couplings $\lambda_\nu$ are allowed compared to the $X$ scalar case.
    Interestingly, the analysis indicates that the $X$ vector case is compatible with slightly larger values of $H_0$ and is compatible with $H_0 > 70$ at the $2\sigma$ level for the region of high masses and low coupling.
    This is due to the contribution of the $X$ particle to $N_{\rm{eff}}$ as explained above and in the main text.
    \item The scenario of $X$ being of scalar boson with $\Delta N_{\rm{eff}}^{\rm{BBN}} = 0$ and $N_{\rm{int}} = 1$. 
    The likelihoods are the same as before.
    The result of the analysis is shown in Figure~\ref{fig:SM_getdist_2}.
    It clearly highlights the non-trivial $1\sigma$ preferred region in the plane $(m_X, \lambda_\nu)$. 
    In particular, this region indicates a slight preference for neutrino-$X$ interactions such that the $X$ boson starts reducing neutrino free-streaming by redshift $z\sim 1000-3500$,  see Eq.~\eqref{eq:GNF_1sigma}.
    \item The result for the scenario which allows a primordial abundance of the $X$ scalar boson and $N_{\rm{int}} = 1$ is shown in Figure~\ref{fig:SM_getdist_3}. 
    Different combinations of $\Delta N_{\rm{eff}}^{\rm{BBN}} \geq  0$, $g_X = (1,3)$ and $N_{\rm{int}} = (1,2,3)$ lead to qualitative same results with only slight quantitative differences, as can be seen from Figure~\ref{fig:background_detail}. 
    The likelihoods are the same as before but now include also a) the Pantheon data set and b) the Pantheon data set together with the $\rm{SH_0ES}$ prior.
    Contrary to what was found in Ref.~\cite{Escudero:2019gvw}, our refined analysis shows that even in the case b), see Table~\ref{tab:cases}, the $H_0$ value predicted by the model can not be increase to the $1\sigma$ $\rm{SH_0ES}$ measured value.
    We find no significant increment in the prediction of the $H_0$ parameter. 
    More details on the quantification of the $H_0$-tension can be found in Section~\ref{sec:results} of the main text.
    \item For the case of a scalar interacting with $N_{\rm int} = 3$ neutrinos and with $\Delta N_{\mathrm{eff}}^{\mathrm{BBN}} = 0$, we show in Figure~\ref{fig:SM_getdist_4} the full correlation of the standard cosmological parameters. 
    The result is compared to $\Lambda$CDM in the same figure. 
    We see that both cosmologies lead to similar correlations in these parameters modulo a small shift on $H_0$ and a multimodal posterior in the $H_0$-$\omega_{\rm CDM}$ plane.
    \end{itemize}


\end{document}